\documentclass[prd,aps,a4paper,superscriptaddress,twocolumn,nofootinbib,floatfix,longbibliography]{revtex4-2}
\pdfoutput=1
\usepackage{graphicx}
\usepackage{color}
\usepackage{dcolumn}
\usepackage{bm}
\usepackage{slashed}
\usepackage{amsmath}
\usepackage{latexsym}
\usepackage{amssymb}
\usepackage{mathrsfs}
\usepackage{placeins}
\usepackage{booktabs}
\usepackage{subfigure}
\usepackage{float}
\usepackage[normalem]{ulem}
\usepackage{threeparttable}
\usepackage{amsfonts}
\usepackage{wrapfig}
\usepackage{autobreak}
\usepackage{mathtools}
\usepackage{multirow}
\usepackage{array}
\usepackage{epstopdf}

\allowdisplaybreaks
\usepackage[linktoc=none]{hyperref}
\hypersetup{colorlinks,			
	linkcolor=blue,
	citecolor=blue}
\allowdisplaybreaks
\begin{document}

\title{
Astrophysically Realistic Secondary Spins Trigger Chaos in Schwarzschild Spacetime and Discernible Gravitational Wave Signatures}

\author{Dan-Dan Yuan}
\email{yuandandan24@mails.ucas.ac.cn}
\affiliation{International Centre for Theoretical Physics Asia-Pacific, University of Chinese Academy of Sciences, 100190 Beijing, China}

\author{Jia-Geng Jiao}
\email{jiaojiageng@ucas.ac.cn}
\thanks{Corresponding author}
\affiliation{International Centre for Theoretical Physics Asia-Pacific, University of Chinese Academy of Sciences, 100190 Beijing, China}

\author{Yu-Qi Lei}
\email{yuqi\_lei@shu.edu.cn}
\thanks{Corresponding author}
\affiliation{Department of Physics, College of Sciences, Shanghai University, 200444 Shanghai, China}

\author{Jun-Xi Shi}
\email{shijunxi23@mails.ucas.ac.cn}
\affiliation{International Centre for Theoretical Physics Asia-Pacific, University of Chinese Academy of Sciences, 100190 Beijing, China}

\author{Jing-Qi Lai}
\email{laijingqi19@mails.ucas.ac.cn}
\affiliation{School of Physical Sciences, University of Chinese Academy of Sciences, Beijing 100049, China}

\author{Caiying Shao}
\email{shaocaiying@ucas.ac.cn}
\thanks{Corresponding author}
\affiliation{School of Physical Sciences, University of Chinese Academy of Sciences, Beijing 100049, China}

\author{Yu Tian}
\email{ytian@ucas.ac.cn}
\thanks{Corresponding author}
\affiliation{School of Physical Sciences, University of Chinese Academy of Sciences, Beijing 100049, China}

\begin{abstract}
Chaos in extreme-mass-ratio inspirals is often thought to require unrealistically large secondary spins, making its astrophysical relevance uncertain. However, we find that chaos persists across the astrophysically realistic spin range for a spinning secondary orbiting a Schwarzschild black hole. This nonintegrable dynamics leaves clear signatures in the emitted gravitational waves. Nearby regular and chaotic trajectories can remain similar in the time domain and retain broadly aligned dominant spectral peaks, yet chaotic signals develop a much less discrete frequency-domain structure with dense inter-peak power. Furthermore, we introduce a local spectral-flatness measure and find it to be several hundred times larger for the chaotic signal than for the neighboring regular signals. Finally, a change in the secondary spin by as little as \(1\%\) of its maximal physically allowed value can drive the system from regular to chaotic motion and produce distinctive detector-level waveforms.
\end{abstract}

\maketitle

\section{Introduction}

Extreme mass ratio inspirals (EMRIs), in which a stellar-mass compact object with mass \(\mu \sim O(1\!-\!100)\,M_\odot\) inspirals into a supermassive black hole with mass \(M \sim 10^{5}\!-\!10^{7}\,M_\odot\), corresponding to a mass ratio \(q \equiv \mu/M \sim 10^{-7}\!-\!10^{-4}\), are among the primary targets of future space-borne gravitational-wave detectors such as LISA \cite{amaro-seoane2017}, Taiji \cite{ruan2020}, and TianQin \cite{luo2016}. Their millihertz gravitational-wave signals can remain in band for months to years, making EMRIs exceptionally powerful probes of the strong-field spacetime geometry of massive black holes and enabling remarkably precise measurements of intrinsic source parameters \cite{babak2017,gair2013}.

At the same time, this high precision makes EMRI waveform modeling especially demanding, since even subleading dynamical effects can accumulate appreciable phase shifts over the long inspiral and leave non-negligible imprints on the observed waveform \cite{barack2009,barack2019,cardenas-avendano2024}. Such effects may arise from astrophysical environments, including nearby companions \cite{bonga2019}, accretion disks \cite{kocsis2011,yunes2011,speri2023,chatterjee2023,duque2025,li2025}, and matter distributions around the central black hole \cite{zhang2019,zhang2020,brito2023}, as well as from finite-size corrections associated with the secondary itself \cite{piovano2025a,skoupy2023b}. Identifying which effects can qualitatively alter the orbital dynamics and how these changes are encoded in the gravitational-wave signal is therefore of direct importance for future EMRI data analysis \cite{lisaconsortiumwaveformworkinggroup2025}.

Among such effects, the spin of the secondary is of particular interest. Beyond the leading geodesic approximation, the smaller compact object is naturally modeled as a pole-dipole body, or equivalently a ``spinning test particle'' governed by the Mathisson-Papapetrou-Dixon (MPD) equations \cite{papapetrou1951,dixon1964,mathisson2010}, whose finite-size effect enters through spin-curvature coupling. Recent progress on generic spinning-secondary EMRIs in Kerr spacetime has made clear that this correction is astrophysically relevant for precision modeling \cite{cui2025a,skoupy2026,cui2026,drummond2026,tan2025}. At linear order in the secondary spin, the dynamics still admits a Carter-like constant and remains integrable \cite{witzany2019,witzany2019a,skoupy2025a}, a property that has recently been exploited to accelerate waveform and flux calculations \cite{skoupy2026,drummond2026}. Once nonlinear spin terms are retained, however, this integrable structure is generally lost. Early studies showed that spin can induce resonances and chaos in black-hole spacetimes \cite{suzuki1997a,kao2005a,hartl2003b,hartl2003c}, but they also suggested that such behavior is mainly associated with unrealistically large spins. This picture was revised significantly by Zelenka \textit{et al.} \cite{zelenka2020a}, who showed that nonintegrable behavior is not confined to unrealistically large spins but can already appear even near the astrophysically allowed upper end of the EMRI spin range. More broadly, understanding whether such spin-induced nonintegrability leaves identifiable gravitational-wave signatures is relevant to the wider phenomenology of nonintegrable EMRIs \cite{lukes-gerakopoulos2021}, where broken hidden symmetries, resonance structures, or environmental perturbations can likewise modify orbital dynamics and waveform morphology \cite{apostolatos2009a,lukes-gerakopoulos2010,destounis2021a,destounis2026,dai2024,hua2026}.

Nevertheless, an important question remains open. Even if chaos can occur for such astrophysically realizable spins, it is not obvious whether it is sufficiently persistent in the relevant phase-space region to matter phenomenologically, nor is it clear whether it leaves distinctive and potentially useful signatures in the corresponding gravitational wave signal.

To address this question, we study the simplest setting in which the effect can be isolated: a spinning secondary moving in Schwarzschild spacetime and governed by the full MPD equations with \emph{non-linearized} spin. Focusing on the astrophysically realistic secondary-spin regime, we analyze the phase-space structure and the associated gravitational-wave signals using the geodesic homoclinic orbit as a reference. We find that chaos is not a fine-tuned feature confined to unrealistically large spins, but persists throughout the realistic spin range considered here. We further show that this nonintegrable dynamics leaves clear imprints on the gravitational-wave signal: nearby regular and chaotic trajectories can remain similar in the time domain and retain aligned dominant spectral peaks, but chaotic signals develop a much denser population of inter-peak components and a markedly less discrete frequency-domain structure. We quantify this behavior with a local spectral-flatness measure and show that the corresponding signature survives detector response. Moreover, even a \(1\%\) change in the secondary spin relative to the astrophysical limit can qualitatively alter the phase-space structure and produce sharply distinguishable detector-level waveforms. Although the characteristic strains remain in the mHz band relevant for space-based detectors and yield comparable overall signal-to-noise ratios, the overlap drops sharply once such a spin change drives the system from regular to chaotic motion. Chaos may therefore leave distinctive detector-level signatures in future space-based gravitational wave observations.

This paper is organized as follows. In Sec.~II, we introduce the dynamical model of spinning particles in Schwarzschild spacetime, discuss the reduced system, and describe the phase-space diagnostics and initial conditions used in this work. In Sec.~III, we present the numerical evidence for chaos in the astrophysically realistic secondary-spin regime through Poincar\'e sections, rotation curves, and Lyapunov indicators. In Sec.~IV, we investigate the corresponding gravitational-wave signatures using numerical kludge waveforms, detector strains, energy spectra, local spectral flatness, and detector-relevant diagnostics such as characteristic strains, signal-to-noise ratios, and overlaps under spin variations. Finally, in Sec.~V we summarize our results and discuss their implications and possible extensions.

\section{Dynamical model of spinning particles in Schwarzschild spacetime}

\subsection{Equations of motion for spinning particles}

On timescales shorter than the radiation-reaction timescale, we model the EMRI as the motion of an extended body in the fixed spacetime background of a static black hole. In the pole-dipole approximation, where only the mass monopole and spin dipole of the secondary object are retained while the quadrupole and higher-order moments are neglected, the motion is described by the MPD equations \cite{papapetrou1951,dixon1964,mathisson2010}:
\begin{align}
\frac{dx^\mu}{d\tau} &= v^\mu, \label{eq:mpd_x} \\
\frac{DP^\mu}{d\tau} &=
-\frac{1}{2} R^\mu{}_{\nu\rho\sigma} v^\nu S^{\rho\sigma}, \label{eq:mpd_p} \\
\frac{DS^{\mu\nu}}{d\tau} &=
P^\mu v^\nu - P^\nu v^\mu. \label{eq:mpd_s}
\end{align}

where \(x^\mu\) denotes the worldline coordinates of the particle, \(\tau\) is the proper time, \(v^\mu\) is the four-velocity, \(P^\mu\) is the four-momentum, and \(S^{\mu\nu}\) is the spin tensor. The right-hand side of Eq.~\eqref{eq:mpd_p} represents the spin-curvature coupling. Setting \(S^{\mu\nu}=0\) recovers geodesic motion for a nonspinning test particle.

Eqs.~\eqref{eq:mpd_x}--\eqref{eq:mpd_s} are not sufficient to determine the evolution of the system because the relation between the four-velocity \(v^\mu\) and the four-momentum \(P^\mu\) is not completely specified within the MPD system. An additional spin supplementary condition (SSC) is therefore needed to fix the representative centroid worldline and thereby close the system. In this work, we adopt the Tulczyjew-Dixon (TD) SSC \cite{tulczyjew1959,dixon1970}:
\begin{equation}
P_\mu S^{\mu\nu} = 0.
\label{eq:td_ssc}
\end{equation}

Under the TD SSC, the dynamical mass and the spin magnitude,
\begin{equation}
\mu^2 = - P_\mu P^\mu,
\label{eq:mu_def}
\end{equation}
\begin{equation}
S^2 = \frac{1}{2} S_{\mu\nu} S^{\mu\nu},
\label{eq:S_def}
\end{equation}
are conserved along the motion \cite{wald1972}. The four-velocity can then be written explicitly as
\begin{equation}
v^\mu = \frac{m}{\mu^2}\left(P^\mu + w^\mu\right),
\label{eq:v_explicit}
\end{equation}
where \(m\equiv -P_\nu v^\nu\) is generally not conserved \cite{semerak1999a}. Using this definition together with the normalization condition \(v^\mu v_\mu=-1\), one finds
\begin{subequations}\label{eq:velocity_explicit}
\begin{align}
m
&=
\frac{\mathcal{A}\mu^2}{\sqrt{\mathcal{A}^2\mu^2-\mathcal{B}S^2}},
\label{eq:m_explicit}\\
w^\mu
&=
\frac{2S^{\mu\nu}R_{\nu\rho\kappa\lambda}P^\rho S^{\kappa\lambda}}{\mathcal{A}},
\label{eq:w_def}\\
\mathcal{A}
&=
4\mu^2 + R_{\alpha\beta\gamma\delta}S^{\alpha\beta}S^{\gamma\delta},
\label{eq:A_def}\\
\mathcal{B}
&=
4 h^{\kappa\eta}
R_{\kappa\iota\lambda\mu} P^\iota S^{\lambda\mu}
R_{\eta\nu\omega\pi} P^\nu S^{\omega\pi},
\label{eq:B_def}\\
h^\kappa{}_\eta
&=
\frac{1}{S^2}S^{\kappa\rho}S_{\eta\rho}.
\label{eq:h_def}
\end{align}
\end{subequations}
Therefore, once the TD SSC is imposed, \(\dot{x}^\mu\), \(\dot{P}^\mu\), and \(\dot{S}^{\mu\nu}\) are uniquely determined, so that the system forms a closed set of ordinary differential equations (ODEs).

Furthermore, if the background spacetime possesses a symmetry generated by a Killing vector \(\xi^\mu\), one can show that the quantity
\begin{equation}
C \equiv \xi_\mu P^\mu - \frac{1}{2}\,\xi_{\mu;\nu} S^{\mu\nu}
\label{eq:killing_constant}
\end{equation}
is conserved along the motion \cite{semerak1999a}. 

\subsection{The Schwarzschild spacetime background}

We consider the Schwarzschild spacetime, which describes a static and spherically symmetric black hole of mass \(M\). In Schwarzschild coordinates, the metric is
\begin{equation}
ds^2=-f(r)\,dt^2+\frac{dr^2}{f(r)}+r^2\left(d\theta^2+\sin^2\theta\,d\phi^2\right),
\label{eq:sch_metric}
\end{equation}
with
\begin{equation}
f(r)=1-\frac{2M}{r}.
\label{eq:sch_f}
\end{equation}
Because the spacetime is static and spherically symmetric, it admits one timelike and three rotational Killing vector fields, given by
\begin{subequations}\label{eq:sch_killing}
\begin{align}
\xi_{(t)} &= \frac{\partial}{\partial t}, \label{eq:sch_killing_t}\\
\xi_{(x)} &= -\sin\phi\,\frac{\partial}{\partial\theta}
            -\cos\phi\cot\theta\,\frac{\partial}{\partial\phi}, \label{eq:sch_killing_x}\\
\xi_{(y)} &= \cos\phi\,\frac{\partial}{\partial\theta}
            -\sin\phi\cot\theta\,\frac{\partial}{\partial\phi}, \label{eq:sch_killing_y}\\
\xi_{(z)} &= \frac{\partial}{\partial\phi}. \label{eq:sch_killing_z}
\end{align}
\end{subequations}

Using Eq.~\eqref{eq:killing_constant}, the conserved quantities associated with the Killing vector fields in Schwarzschild spacetime can be written explicitly as \cite{ehlers1977}
\begin{subequations}\label{eq:sch_constants}
\begin{align}
E &\equiv -C\!\left(\xi_{(t)}\right)
   = -P_t-\frac{M}{r^2}S^{tr}, \label{eq:sch_E}\\
J_x &\equiv C\!\left(\xi_{(x)}\right) \notag\\
    &= -\sin\phi\,P_\theta-\cos\phi\cot\theta\,P_\phi
    +r^2\cos\phi\sin^2\theta\,S^{\theta\phi} \notag\\
    &\quad +r\sin\phi\,S^{\theta r}
    +r\cos\phi\sin\theta\cos\theta\,S^{\phi r}, \label{eq:sch_Jx}\\
J_y &\equiv C\!\left(\xi_{(y)}\right) \notag\\
    &= \cos\phi\,P_\theta-\sin\phi\cot\theta\,P_\phi
    +r^2\sin\phi\sin^2\theta\,S^{\theta\phi} \notag\\
    &\quad -r\cos\phi\,S^{\theta r}
    +r\sin\phi\sin\theta\cos\theta\,S^{\phi r}, \label{eq:sch_Jy}\\
J_z &\equiv C\!\left(\xi_{(z)}\right)
    = P_\phi-r\sin^2\theta\,S^{\phi r}
    -r^2\sin\theta\cos\theta\,S^{\phi\theta}. \label{eq:sch_Jz}
\end{align}
\end{subequations}

Here \(E\) is the conserved energy, while \((J_x,J_y,J_z)\) are the conserved components of the total angular momentum, whose magnitude is defined by
\begin{equation}
J^2 = J_x^2 + J_y^2 + J_z^2.
\label{eq:J2_def}
\end{equation}

\subsection{Reduction of the system}

To set up the orbital families studied in this work, it is useful to briefly distinguish between the geodesic Schwarzschild dynamics and the spinning-particle dynamics. For a nonspinning massive particle in Schwarzschild spacetime, the motion is integrable and the radial equation can be reduced to a one-dimensional problem with a genuine effective potential. By contrast, for a spinning particle described by the full MPD equations with the TD supplementary condition, the system is in general non-integrable, even though several conserved quantities remain available \cite{suzuki1997a,tanaka1996}. This distinction is essential for understanding both the choice of initial conditions and the origin of the chaotic structures analyzed later. 

Owing to the spherical symmetry, the orbit of a nonspinning massive particle in Schwarzschild spacetime can always be confined to a fixed plane, which we choose as the equatorial plane \(\theta=\pi/2\) without loss of generality. 
Using the conserved energy \(E\), orbital angular momentum \(L\), and the normalization of the four-velocity, the geodesic equations can be reduced to the radial equation
\begin{equation}
\mu^2\left(\frac{dr}{d\tau}\right)^2
+
V_{\rm geo}^2(r;L)
=
E^2,
\label{eq:geo_radial_V}
\end{equation}
where the geodesic effective potential is
\begin{equation}
V_{\rm geo}^2(r;L)
=
\left(1-\frac{2M}{r}\right)
\left(\mu^2+\frac{L^2}{r^2}\right).
\label{eq:Vgeo}
\end{equation}
The allowed region of motion is then determined by \(E^2 \ge V_{\rm geo}^2(r;L)\). 
In this sense, the geodesic effective potential provides a genuine one-dimensional description of the orbital structure: its extrema correspond to stable and unstable circular orbits, while the separatrix between bound and plunging motion is given by a homoclinic orbit asymptotic to the unstable circular orbit \cite{levin2009a,kao2005a}.

In the spinning case, however, the situation is qualitatively different. Due to the spin-curvature coupling, the motion is generally non-integrable, even though the conserved quantities \(E\), \(J_z\), and \(J^2\) remain available. Nevertheless, the Schwarzschild symmetries still allow a useful reduction of the dynamics to a two degrees-of-freedom (DoF) system \cite{witzany2019}. In particular, by exploiting the spherical symmetry of Schwarzschild spacetime, one may choose the coordinate system such that the total angular momentum is aligned with the \(z\)-axis,
\begin{equation}
J_x=J_y=0, \qquad J_z=J,
\label{eq:Jalign_reduced}
\end{equation}
which implies
\begin{equation}
P_\theta=-\cot\theta\,P_\phi+r^2S^{\theta\phi}=0.
\label{eq:Ptheta_zero}
\end{equation}

In this reduced setting, although no effective potential exists on the \(r\)-\(\theta\) plane analogous to the geodesic case, one can still characterize the boundary of the allowed motion through two branches of an ``effective potential'' defined by the turning-point conditions \(P_r=P_\theta=0\) \cite{zelenka2020a,suzuki1997a}:
\begin{subequations}\label{eq:Veff_full}
\begin{align}
E &= V_{\mathrm{eff}}^{(\pm)}(r,\theta;J,S), 
\label{eq:Veff_main}\\
V_{\mathrm{eff}}^{(\pm)}
&=
\mu\Biggl[
\sqrt{f}\,\cosh X^{(\pm)}
+\frac{M\sinh X^{(\pm)}}{\sqrt{f}\,r\cosh X^{(\pm)}}
\notag\\*
&\qquad\qquad\times
\left(
\frac{J\sin\theta}{\mu r}-\sinh X^{(\pm)}
\right)
\Biggr],
\label{eq:Veff_def}\\
\sinh X^{(\pm)}
&= \alpha \pm \sqrt{\alpha^2-\beta},
\label{eq:Xpm_def}\\
D
&= \mu^2 r^2 - S^2 f,
\label{eq:D_def}\\
\alpha
&= \frac{\mu J r\sin\theta}{D},
\label{eq:alpha_def}\\
\beta
&=
\frac{(J^2-S^2)f+\dfrac{2M}{r}J^2\sin^2\theta}{D}.
\label{eq:beta_def}
\end{align}
\end{subequations}
Here \(f(r)=1-2M/r\). In our convention, the two branches \(V_{\mathrm{eff}}^{(\pm)}\) may be interpreted as corresponding to spin-aligned and spin-anti-aligned configurations relative to the orbital angular momentum. For the astrophysically realistic small spins considered here, the motion remains very close to the equatorial plane. The reason is that the reality condition of Eq.~\eqref{eq:Xpm_def} restricts the allowed polar angle: in the large-\(r\) limit one finds \(\cos\theta \lesssim S/J_z\), and the bound is typically even slightly stronger at finite \(r\). Since \(S/J_z\) is very small in the regime of interest, \(\theta\) can deviate from \(\pi/2\) only slightly, so the particle is confined to a narrow wedge around the equatorial plane. In the limit \(S\to0\), this wedge shrinks to the equatorial plane itself, recovering the geodesic case.

While \(V_{\mathrm{eff}}^{(\pm)}\) is not a potential in the usual dynamical sense, it still provides a useful orbital picture of the reduced dynamics of the spinning particle: it delineates the allowed and forbidden regions of motion, reveals the appearance of minima and saddle points, and offers a convenient way to understand how the geodesic separatrix and homoclinic reference structure are deformed once spin is introduced. It therefore serves not only as a practical tool for selecting bound initial conditions but also as a guide to identifying the phase-space region in which chaotic motion is most likely to emerge.

\subsection{Phase space diagnostics}

A powerful tool for visualizing the phase space structure of a conservative system is the Poincar\'e surface of section. For a system with 2 DoF, bounded motion is confined to a three-dimensional hypersurface of constant energy in the four-dimensional phase space. Choosing a two-dimensional surface transverse to the Hamiltonian flow, one obtains a Poincar\'e section, on which successive intersections of a trajectory define a discrete-time map.

For an integrable system, the Liouville-Arnold theorem implies that bounded motion lies on two-dimensional invariant tori characterized by two fundamental frequencies, \(\omega_1\) and \(\omega_2\). If the ratio \(\omega=\omega_1/\omega_2\) is irrational, the motion is quasiperiodic and the corresponding intersections on the Poincar\'e section lie on a closed invariant curve. If the ratio is rational, \(\omega=q/p\), the torus is resonant and a single trajectory gives rise to a finite set of \(p\) periodic points on the section \cite{zelenka2020a}.

When integrability is weakly broken, the transition to nonintegrability is governed qualitatively by the KAM \cite{moser1962,givental2009,chierchia2008} and Poincar\'e-Birkhoff theorems \cite{poincare1912,birkhoff1913}. The KAM theorem states that most non-resonant invariant tori survive under sufficiently small perturbations, appearing on the section as slightly deformed invariant curves \cite{moser1962,givental2009,chierchia2008}. By contrast, resonant tori are structurally fragile: each resonant torus breaks into a Birkhoff chain consisting of an even number of periodic points on the section, with elliptic and hyperbolic points alternating along the chain \cite{poincare1912,birkhoff1913}. Consequently, the perturbed Poincar\'e section typically exhibits a central fixed point surrounded by nested KAM curves, island chains around elliptic periodic points, and thin chaotic layers developing near hyperbolic structures. The same mechanism recurs at higher-order resonances, leading to a hierarchical phase-space structure with progressively finer island chains and chaotic layers.

In the present problem, however, our main interest is not in these secondary hyperbolic structures generated around higher-order resonances, but in the primary hyperbolic skeleton already present in the unperturbed Schwarzschild limit. This zeroth-order hyperbolic structure is the unstable circular orbit underlying the geodesic separatrix, which appears as a hyperbolic fixed point on the Poincar\'e section. In the integrable limit, its stable and unstable manifolds coincide to form the homoclinic separatrix; once integrability is broken by spin-curvature coupling, chaotic motion is expected to emerge first in the neighborhood of this homoclinic structure \cite{kao2005a,guckenheimer1983}.

A useful quantitative diagnostic of the structures on a Poincar\'e section is the rotation number. For successive intersection points \(w_n\), one defines \(R_n=w_n-u_0\) relative to a fixed reference point \(u_0\), and lets \(\vartheta_n\) be the angle between \(R_n\) and \(R_{n+1}\). The finite-time rotation number is then
\begin{equation}
\nu_\vartheta=\frac{1}{2\pi N}\sum_{i=1}^{N}\vartheta_i ,
\end{equation}
which, as \(N\to\infty\), approaches the frequency ratio \(\nu_\vartheta=\omega_1/\omega_2\), allowing the rotation number to serve as a convenient indicator of resonances \cite{efthymiopoulos1999}. For regular motion, \(\nu_\vartheta\) varies smoothly across neighboring invariant curves; resonances appear as rational plateaus in the rotation curve, while irregular variations are indicative of chaotic layers \cite{contopoulos2002,lukes-gerakopoulos2010,destounis2020}. 

\subsection{Initial conditions}

\begin{figure}[htbp]
    \centering
    \includegraphics[width=\columnwidth]{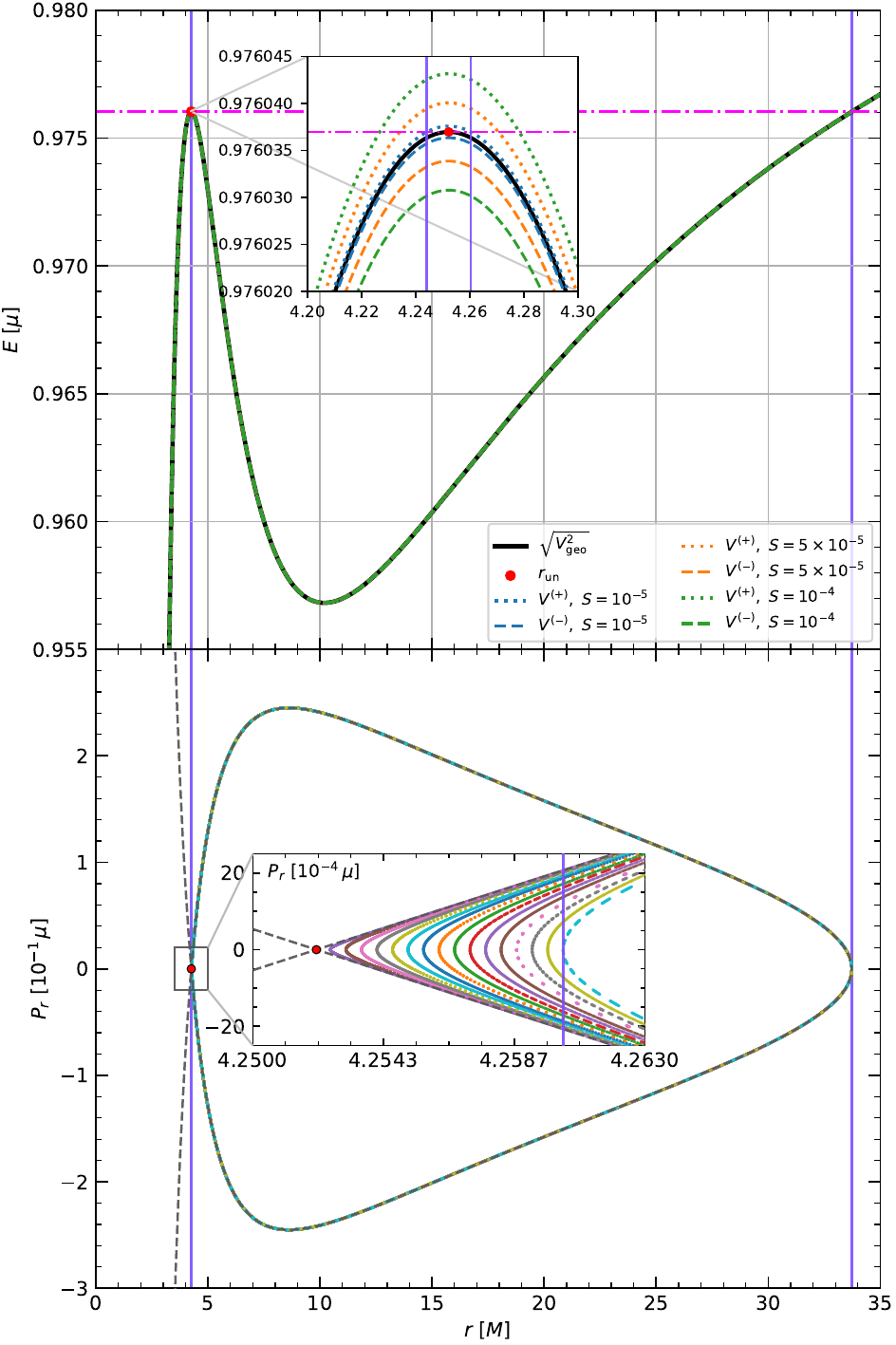}
    \caption{Top panel: The effective potential on the equatorial plane for \(J_z=3.8\mu M\). The full black line corresponds to \(S=0\). For \(S=10^{-5}\), \(5\times10^{-5}\), and \(10^{-4}\), the dotted and dashed colored lines denote \(V_{\rm eff}^{(+)}\) and \(V_{\rm eff}^{(-)}\), respectively. The magenta dash-and-dot line marks the geodesic homoclinic energy \(E_{\rm h}\). The inset enlarges the neighborhood of the potential peak. The red point marks \(r_{\rm un}\), the hyperbolic fixed point associated with the homoclinic orbit. Bottom panel: Poincar\'e sections on the \((r,P_r)\) plane for \(S=10^{-5}\), \(J_z=3.8\mu M\) and \(E=E_{\rm h}\), obtained from 21 initial conditions with \(r_0\) uniformly sampled from \(4.25M\) to \(4.260289M\). The gray dashed line shows the geodesic homoclinic orbit. The blue-purple vertical lines connecting both panels indicate the bounds for possible initial \(r_0\) with \(P_r=0\) in the Poincar\'e section and the red point marks the position of \(r_{\rm un}\) on the \((r,P_r)\) plane.}
    \label{fig:Veff}
\end{figure}
Since our motivation is astrophysical EMRIs, the spin of the secondary should be restricted to astrophysically realistic values. For a stellar-mass black hole, the astrophysical spin angular momentum is bounded by \(\mu^2\), while for a neutron star, the mass-shedding limit gives the more restrictive bound \(\lesssim 0.6\,\mu^2\) \cite{hartl2003b}. Therefore, the dimensionless spin parameter entering the MPD equations satisfies
\begin{equation}
\frac{S}{\mu M}\lesssim \frac{\mu}{M},
\end{equation}
with the more restrictive bound \(S\lesssim 0.6\,\mu/M\) for a neutron-star secondary \cite{cook1994}. For the mass ratio considered in this work, \(\mu/M=10^{-4}\), this implies that astrophysically realistic spins are confined to \(S\lesssim10^{-4}\). In the following, we restrict attention to the astrophysically relevant spin range allowed in EMRIs.
\begin{table}[htbp]
\centering
\caption{The numerical values of the orbital angular momentum \(L\), the unstable point \(r_{\rm un}\), the maximum point \(r_{\max}\), the stable point \(r_{\rm st}\), and the homoclinic energy \(E_{\rm h}\) for the reference homoclinic orbit in Schwarzschild spacetime.}
\label{tab:homo_baseline}
\renewcommand{\arraystretch}{1.25}
\begin{ruledtabular}
\begin{tabular}{ccccc}
$L~[\mu M]$ & $r_{\rm un}~[M]$ & $r_{\max}~[M]$ & $r_{\rm st}~[M]$ & $E_{\rm h}~[\mu]$ \\
\hline
3.8 & 4.252105 & 33.732794 & 10.187895 & 0.976036965 \\
\end{tabular}
\end{ruledtabular}
\end{table}

As discussed in the previous subsection, for Schwarzschild geodesics, the homoclinic orbit is associated with the unstable circular orbit and provides the natural phase space neighborhood relevant to the onset of chaos when integrability is weakly broken by spin-curvature coupling. Such geodesic homoclinic orbits exist for \(L_{\rm ISCO}<L<L_{\rm IBCO}\), i.e., \(2\sqrt{3}\,\mu M<L<4\,\mu M\) in Schwarzschild spacetime, where \(L_{\rm ISCO}\) and \(L_{\rm IBCO}\) denote the angular momenta of the innermost stable circular orbit (ISCO) and the innermost bound circular orbit (IBCO), respectively \cite{levin2009a}. In this work, we choose the representative homoclinic baseline with \(L = 3.8\,\mu M\), whose characteristic parameters are listed in Table~\ref{tab:homo_baseline}. In the spinning case, the relevant conserved quantity is the total angular momentum \(J\), rather than the geodesic orbital angular momentum \(L\). Here we set \(J=L\) to use the geodesic homoclinic orbit as the reference configuration so that, for the astrophysically realistic spin range considered in this work, the spin-curvature coupling can be treated as a weak perturbation and the resulting trajectories can be viewed as deformations of the geodesic homoclinic neighborhood.

For the numerical construction of the Poincar\'e sections, we fix \((E,J_z,S)\) and choose the initial radius \(r_0\) on the line \(P_r=0\) within the range of admissible initial conditions allowed by the constraints. We take the equatorial plane \(\theta=\pi/2\) with \(P_\theta\ge 0\) as the surface of section and use \((r,P_r)\) as section coordinates, which is a commonly used choice in black hole spacetimes \cite{seyrich2012b,zelenka2020a}.

Figure~\ref{fig:Veff} illustrates the astrophysical picture underlying our choice of initial conditions. In the top panel, for fixed \(J_z=3.8\mu M\), the two branches \(V_{\rm eff}^{(+)}\) and \(V_{\rm eff}^{(-)}\) delimit a narrow allowed band in the \((r,E)\) plane. For a given energy and with \(P_r=0\), the admissible initial radii are given by the intersection of the energy level with this band, as indicated by the blue-violet solid lines in the inset (see Appendix~A for details). For the astrophysically realistic spins (\(S\le 10^{-4}\)) considered here, if we choose \(E=E_{\rm h}\), the allowed band of motion is only weakly deformed from the geodesic one and remains close to the geodesic homoclinic neighborhood.

The bottom panel shows the corresponding Poincar\'e sections for \(S=10^{-5}\) at \(E=E_{\rm h}\). The gray dashed curve represents the geodesic homoclinic orbit, while the red point marks \(r_{\rm un}\), the hyperbolic fixed point associated with it. Initial conditions with \(r_0<r_{\rm un}\) lie on the plunging side of the weakly deformed separatrix and therefore correspond to rapidly plunging trajectories. Such orbits intersect the surface of the section only a few times before falling into the black hole, leaving at most a few scattered points rather than persistent phase-space structures. By contrast, initial conditions from the narrow admissible interval outside \(r_{\rm un}\) correspond to bound orbits and form invariant tori on the Poincar\'e section. In this sense, although spin destroys the exact geodesic separatrix, for the astrophysically realistic spin range studied here, the geodesic homoclinic orbit remains an accurate reference for selecting the relevant initial conditions.

\section{Chaos for Astrophysically Realistic Secondary Spins}

\subsection{Chaotic Signatures in Phase Space}

\begin{figure*}[htbp]
    \centering
    \includegraphics[width=\textwidth]{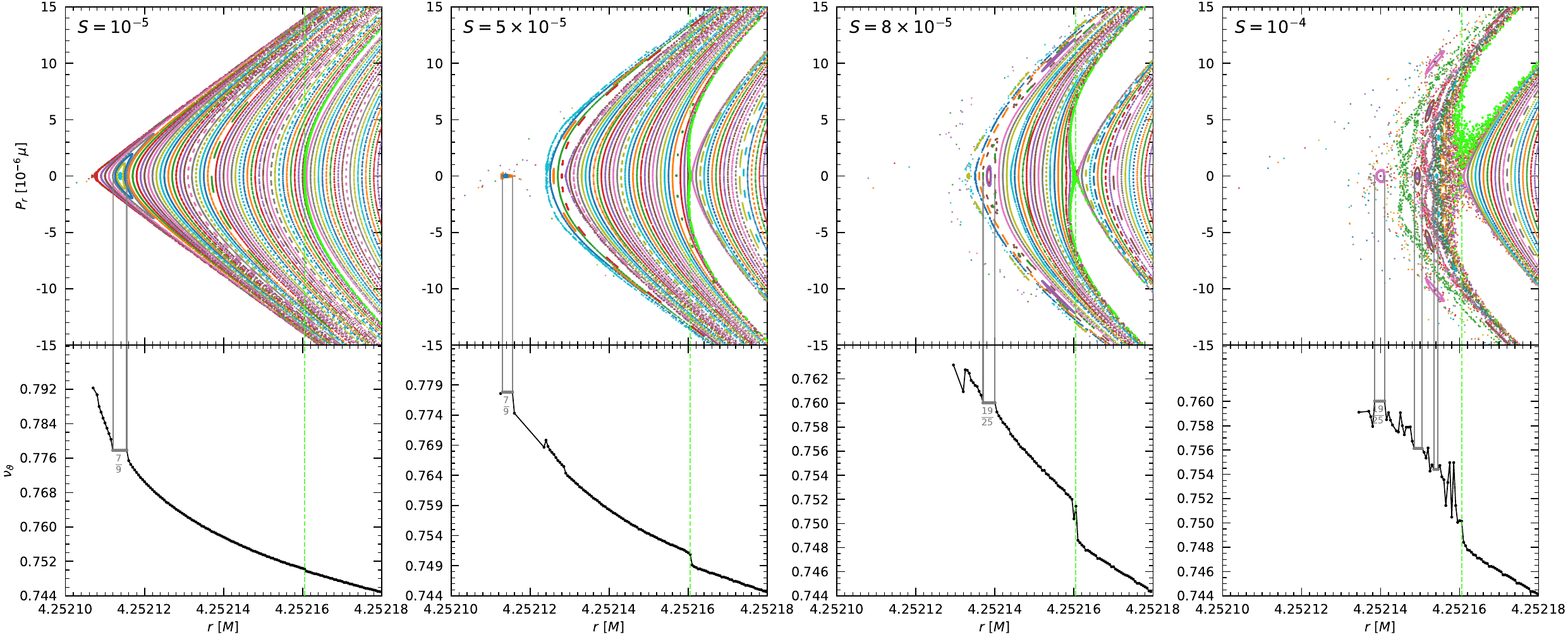}
    \caption{Top panels: Poincar\'e surfaces of section. Bottom panels: corresponding rotation curves. From left to right, \(S=10^{-5}\mu M\), \(5\times10^{-5}\mu M\), \(8\times10^{-5}\mu M\), and \(10^{-4}\mu M\). In all panels, \(E=E_{h}\), \(J_z=3.8\mu M\), and initial \(P_r=0\). The initial radius is sampled uniformly in \(r_0\in[4.252105M,\,4.252180M]\) with step size \(10^{-6}M\). Gray lines mark the dominant resonant islands, and the lime-green lines indicate the separatrix.
    }
    \label{fig:Poincare}
\end{figure*}
We now present the numerical results for the phase-space structure in the neighborhood of the homoclinic orbits identified in Sec.~II E. Figure~\ref{fig:Poincare} shows the Poincar\'e section and the corresponding rotation curves for four values of the secondary spin, ranging from \(S=10^{-5}\mu M\) to \(10^{-4}\mu M\), while keeping the other parameters fixed at \(E=E_{h}\), \(J_z=3.8\mu M\), and \(P_r=0\).

We first consider the top panels of Fig.~\ref{fig:Poincare}. From left to right, as the spin increases, all four sections exhibit the standard structures discussed in Sec.~II D, namely surviving invariant curves and resonant islands. For the smallest spin, \(S=10^{-5}\mu M\), the section is still dominated by regular invariant tori, with only very weak signs of irregularity in the region closer to the black hole. As the spin increases to \(S=5\times10^{-5}\mu M\) and \(8\times10^{-5}\mu M\), the phase space structure near the geodesic homoclinic orbit becomes visibly deformed: some orbits that previously lay on invariant curves now plunge into the black hole, indicating that the homoclinic separatrix of the geodesic baseline is progressively distorted by the spin-curvature coupling. At the same time, increasingly evident chaotic layers develop around the inner region of the section. For the largest spin considered here, \(S=10^{-4}\mu M\), the chaotic layer becomes prominent in the region closer to the black hole, where scattered points occupy an extended area of the section, while the phase-space structure farther from the black hole remains comparatively regular and still contains recognizable resonant islands. This behavior is qualitatively consistent with the picture emphasized by Suzuki and Maeda, namely that the sea of chaos spreads as the spin increases, and it is also in agreement with the result of Zelenka \textit{et al.} for the case \(S=10^{-4}\mu M\), where typical signatures of a non-integrable system already appear for EMRI-relevant spin values \cite{suzuki1997a,zelenka2020a}.

The rotation curves shown in the bottom panels of Fig.~\ref{fig:Poincare} correspond well to the behavior described in Sec.~II D: the rotation number forms plateaus when passing through secondary islands of stability, varies monotonically along regular KAM curves, and fluctuates strongly in chaotic regions. In the cases \(S=5\times10^{-5}\mu M\) and \(S=8\times10^{-5}\mu M\), one can already see pronounced irregular variations in the rotation curve, which match well with the chaotic layers visible in the corresponding Poincar\'e sections. For \(S=10^{-5}\mu M\), the rotation curve is mostly smooth, but a slight loss of smoothness can still be seen near the left edge around \(r\simeq4.252108M\); this corresponds to the very weak chaotic-layer-like feature in the Poincar\'e section and suggests that chaos may already be present there, although only marginally. This point will be further confirmed by additional chaos indicators in the following subsections. For \(S=10^{-4}\mu M\), the rotation curve displays much stronger fluctuations over a wider interval, reflecting the substantial growth of the chaotic layer in phase space.

A further remarkable feature in Fig.~\ref{fig:Poincare} is the sharp jump of the rotation curve near \(\nu_\theta=0.75\), which appears in all four cases and becomes increasingly pronounced as the spin increases. For the larger spin values, \(S=8\times10^{-5}\mu M\) and \(S=10^{-4}\mu M\), clear signatures of chaos develop in its vicinity. In the Poincar\'e sections, this region corresponds to the orbits highlighted by the lime-green lines, which in fact coincide with the separatrix discussed by Zelenka \textit{et al.} \cite{zelenka2020a} in the neighborhood of a resonance, i.e. the constant Hamiltonian curve separating the phase space regions of oscillation and libration. In the present case, it is associated with the \(3/4\) resonance, where a hyperbolic fixed point appears on the Poincar\'e section. As described in Sec.~II D, this is precisely the scenario predicted by the Poincar\'e-Birkhoff theorem: under a weak perturbation, a resonant torus is broken and replaced by alternating stable and unstable periodic points. The hyperbolic fixed point generated in this way gives rise to a new separatrix, which marks the boundary between oscillatory and librational motion and provides the local onset of chaos \cite{poincare1912,birkhoff1913}. Therefore, the onset of chaos seen in Fig.~\ref{fig:Poincare} is directly tied not only to the deformation of the original geodesic phase-space structure but also to the emergence of a new hyperbolic fixed point and its associated homoclinic separatrix near the \(3/4\) resonance.

\subsection{Lyapunov Exponents and Fast Lyapunov Indicators}

To complement the phase space diagnostics of the previous subsection, we now introduce two quantitative chaos indicators, namely the Lyapunov exponent and the fast Lyapunov indicator. In dynamical systems, Lyapunov exponents (LE) quantify the average exponential rate at which initially nearby trajectories separate in phase space. From a computational point of view, the maximum LE can be obtained either from the variational equations or, more conveniently, from the evolution of a small deviation between two nearby trajectories. In curved spacetime, however, the classical definitions based on coordinate time and coordinate distance are generally not invariant under coordinate transformations and may therefore fail to provide a physically meaningful characterization of chaos. Following the coordinate invariant formulation of Lyapunov indicators developed by Wu \textit{et al.} \cite{wu2003a,wu2006a} and its subsequent application to the dynamics of spinning secondaries by Han \cite{han2008a} and Hartl \cite{hartl2003b}, we adopt the two nearby trajectories method in the present work. This construction avoids the derivation and integration of the full variational equations, which are particularly cumbersome and computationally expensive for the MPD dynamics, while still retaining the essential information of the maximum LE. In practice, it provides an efficient and sufficiently accurate way to identify chaotic regions and is therefore well suited for the present problem.

More specifically, let \(x^\mu(\tau)\) denote the reference orbit and \(\tilde{x}^\mu(\tau)\) a nearby shadow orbit, both parameterized by the proper time \(\tau\) of the reference particle. Their deviation vector is defined as
\begin{equation}
\Delta x^\mu(\tau)=\tilde{x}^\mu(\tau)-x^\mu(\tau).
\end{equation}
Since the spinning secondary follows a non-geodesic trajectory, the separation should be measured in the local rest space of the reference orbit. Introducing the projection tensor
\begin{equation}
h_{\mu\nu}=g_{\mu\nu}+v_\mu v_\nu,
\end{equation}
the corresponding proper distance is
\begin{equation}
d(\tau)=\sqrt{\left|h_{\mu\nu}\Delta x^\mu \Delta x^\nu\right|}.
\end{equation}
The Lyapunov exponent is then defined as follows:
\begin{equation}
\lambda=\lim_{\tau\to\infty}\frac{1}{\tau}\ln\frac{d(\tau)}{d(0)},
\end{equation}
while in numerical calculations, we evaluate its finite-time version as
\begin{equation}
\lambda(\tau)=\frac{1}{\tau}\sum_{i=1}^{N(\tau)}\ln\frac{d_i}{d_0},
\end{equation}
where \(d_0\) is the prescribed initial separation and \(d_i\) is the separation immediately before the \(i\)-th renormalization. Such a renormalization is necessary because the distance between the reference orbit and the shadow orbit may grow rapidly and soon saturate at the finite scale of the bounded chaotic region, thereby obscuring the true exponential divergence.

\begin{figure*}[htbp]
    \centering
    \includegraphics[width=\textwidth]{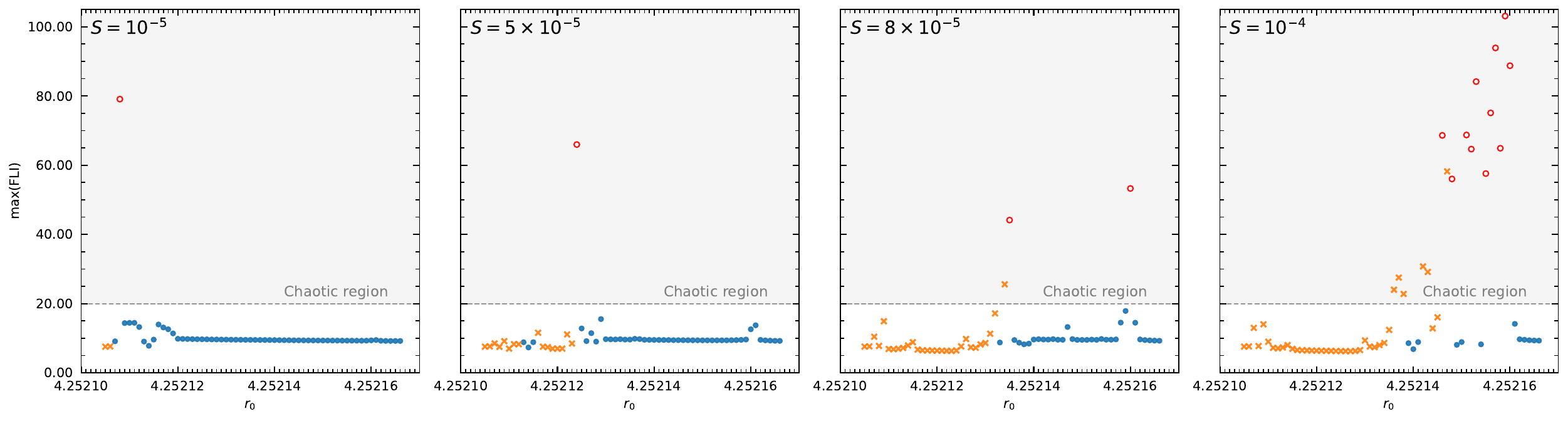}
    \caption{Maximum FLI corresponding to the Poincar\'e sections shown in Fig.~\ref{fig:Poincare}. From left to right, the spin values are \(S=10^{-5}\mu M\), \(5\times10^{-5}\mu M\), \(8\times10^{-5}\mu M\), and \(10^{-4}\mu M\). In each panel, the initial radius is sampled uniformly from \(r_0=4.252105M\) using 62 equidistant values with the spacing \(10^{-6}M\). Orange crosses denote plunging trajectories, blue filled circles regular bound trajectories, and red open circles chaotic trajectories. The gray shaded region marks the criterion adopted here for identifying chaotic motion.
    }
    \label{fig:MaxFLI}
\end{figure*}

Since the finite-time LE usually requires a rather long integration time to approach a stable value, we also introduce the fast Lyapunov indicator (FLI), which provides a more efficient diagnostic of chaos \cite{wu2003a}. In the present two-nearby-trajectories framework, the FLI is constructed from the same proper separation \(d(\tau)\) as
\begin{equation}
\mathrm{FLI}(\tau)=\log_{10}\frac{d(\tau)}{d_0}.
\end{equation}

FLI likewise requires renormalization to avoid saturation for chaotic trajectories. If \(k\) denotes the number of renormalization steps up to time \(\tau\), the renormalized FLI is given by
\begin{equation}
\mathrm{FLI}_k(\tau)=-k\bigl(1+\log_{10}d_0\bigr)+\log_{10}\frac{d(\tau)}{d_0}.
\end{equation}
With this prescription, the FLI retains its sensitivity to chaotic divergence while avoiding saturation and is therefore particularly suitable for efficiently scanning chaotic regions, whereas the LE serves as a complementary long term consistency check.

In our numerical implementation, the scan over initial conditions is sufficiently fine that quadruple-precision floating-point variables are required in order to reliably resolve the divergence between nearby trajectories. After extensive numerical experiments, we set the renormalization distance to \(d_0=10^{-6}\) and perturb the initial radial coordinate by a small offset \(\delta r=10^{-16}M\). For trajectories integrated up to \(\tau=10^{6}M\), we adopt \(\mathrm{FLI}=20\) as the threshold separating chaotic from regular motion, whereas for trajectories integrated only up to \(\tau=10^{5}M\), \(\mathrm{FLI}=10\) provides a suitable criterion.

The resulting maximum FLI values are shown in Fig.~\ref{fig:MaxFLI}, which provides an important complement to, and cross-check of, the Poincar\'e-section analysis in Fig.~\ref{fig:Poincare}. In particular, for the smallest spin \(S=10^{-5}\mu M\), the orbit with \(r_0=4.252108M\), which already appeared suspicious due to the weak irregularity in the corresponding Poincar\'e section and rotation curve, is now clearly identified as chaotic. This confirms that the left edge of the phase-space structure in Fig.~\ref{fig:Poincare} already contains a thin chaotic layer even for astrophysically realistic spins as small as \(10^{-5}\mu M\). As the spin increases, two trends become increasingly evident. First, more and more trajectories plunge into the black hole after sufficiently long integration times; these are marked by orange crosses in Fig.~\ref{fig:MaxFLI}. Second, the number of trajectories classified as chaotic also increases steadily, indicating that the chaotic region broadens as the spin-curvature coupling becomes stronger. In addition, some trajectories display a clear tendency toward chaotic growth before eventually plunging, although their long-time bounded evolution can no longer be tracked further. This suggests that the actual influence of chaos may be even more extensive than what can be inferred solely from the surviving bound orbits.
\begin{figure}[htbp]
    \centering
    \includegraphics[width=\columnwidth]{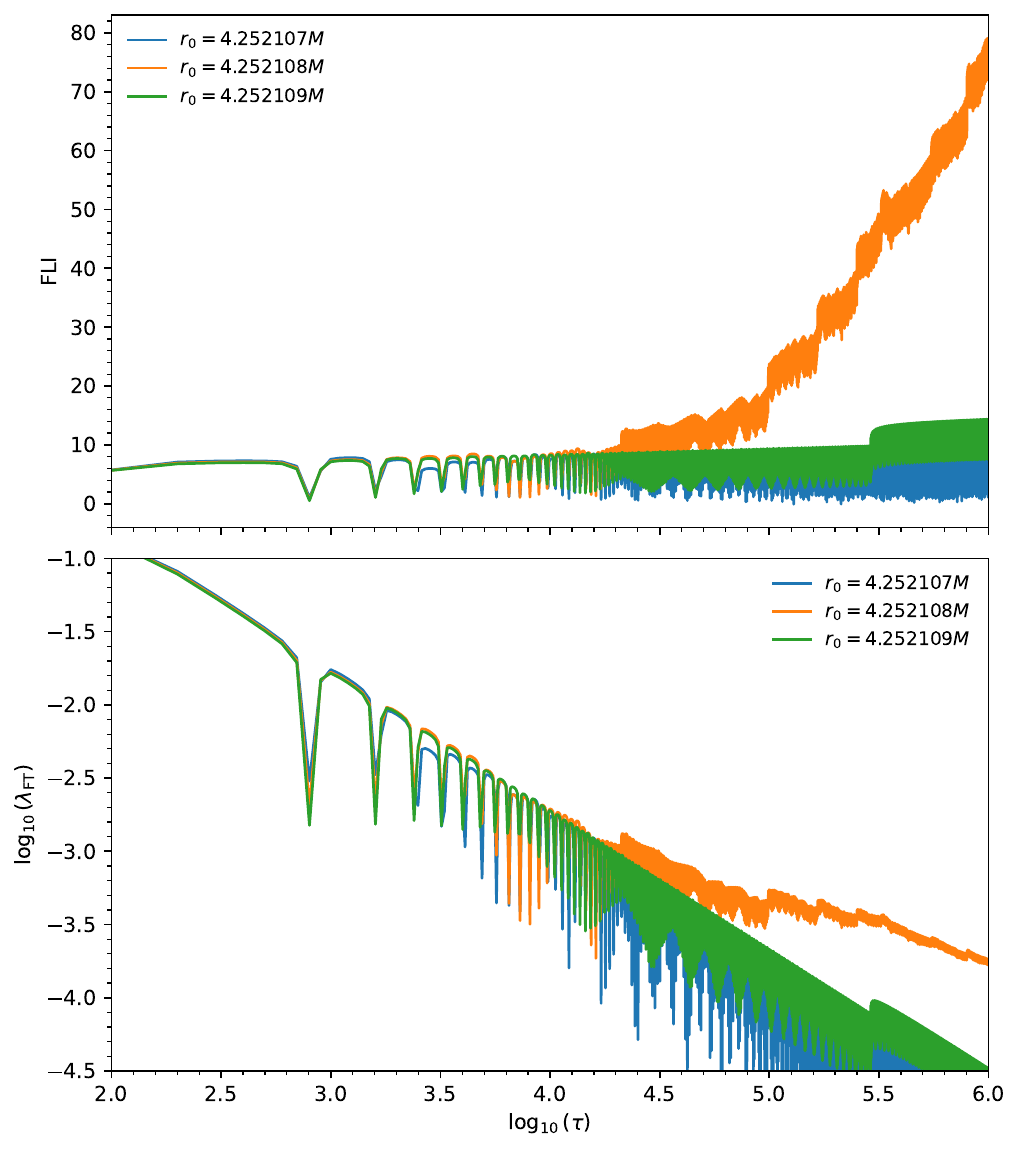}
    \caption{Time evolution of the FLI (top) and LE (bottom) for \(S=10^{-5}\mu M\) and three representative initial radii, \(r_0=4.252107M\), \(4.252108M\), and \(4.252109M\). The orbit with \(r_0=4.252108M\) is chaotic, while the other two remain regular. To ensure convergence of the LE, all trajectories are integrated up to \(\tau=10^{6}M\).
    }
    \label{fig:FLI(t)}
\end{figure}

A more detailed illustration is given in Fig.~\ref{fig:FLI(t)}, which shows the time evolution of the FLI and LE for \(S=10^{-5}\mu M\) and three representative initial radii. The orbit with \(r_0=4.252108M\) exhibits exponential divergence after about \(\tau\sim10^{4}M\), as indicated by the rapid growth of the FLI. By contrast, the FLI of the regular orbits remains at comparatively low values throughout the evolution. The sharp downward spikes visible in the regular curves are caused by the renormalization procedure, which slightly displaces the numerical orbit away from the constrained phase-space surface; however, this artifact does not affect the overall diagnosis of regularity versus chaos. The bottom panel shows that the LE is fully consistent with the behavior indicated by the FLI. At late times, the LE of the regular orbits continues to decrease toward zero, whereas the chaotic orbit settles to a positive value.

\begin{figure}[htbp]
    \centering
    \includegraphics[width=\columnwidth]{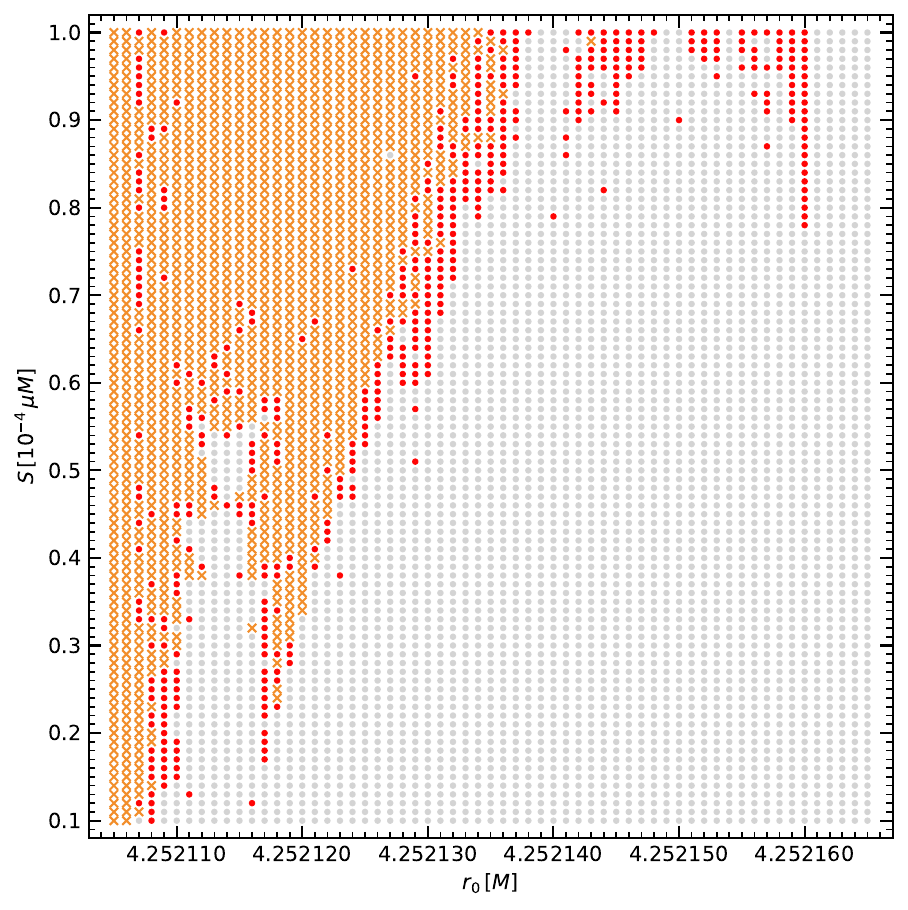}
    \caption{Parameter scan in the \((r_0,S)\) plane. The spin magnitude is sampled from \(S=10^{-5}\mu M\) to \(10^{-4}\mu M\) with step size \(10^{-6}\mu M\), while the initial radius is sampled from \(r_0=4.252105M\) to \(4.252165M\) with step size \(10^{-6}M\), giving a total of 5551 parameter points. All trajectories are integrated up to \(\tau=10^{5}M\). Orange crosses denote plunging orbits, red dots chaotic orbits, and gray dots regular orbits.
    }
    \label{fig:FLI_2D}
\end{figure}

A broader view is provided by the two-dimensional parameter scan in Fig.~\ref{fig:FLI_2D}. Overall, as the spin increases, both plunging and chaotic trajectories become more numerous. However, this dependence is not simply linear: the boundaries between regular, chaotic, and plunging motion remain highly irregular in the \((r_0,S)\) plane, indicating a sensitive dependence on both the spin magnitude and the initial radius. Some of the additional chaotic points appearing in Fig.~\ref{fig:FLI_2D}, compared with Fig.~\ref{fig:MaxFLI}, correspond to trajectories that plunge only on longer timescales. Before plunging, their FLI and LE exhibit the same exponentially diverging behavior as the chaotic bound orbits discussed above. We do not attempt a more detailed classification of such trajectories here.

In summary, by combining the Poincar\'e sections, rotation curves, and Lyapunov indicators, we demonstrate, to the best of our knowledge, for the first time that chaotic motion persists even for secondary spins as small as \(S=10^{-5}\mu M\), corresponding to only \(10\%\) of the maximal physically allowed spin. This shows that, for the EMRI system considered here, namely a spinning secondary moving in a Schwarzschild black hole, chaos is not confined to near-extremal or unrealistically large spins, but is already a generic feature within the astrophysically realistic spin range considered here.

\section{Gravitational Wave Signatures of Chaotic Orbits}

\subsection{Numerical kludge waveforms and detector strains}

Having established in Sec.~III that chaotic motion already occurs for astrophysically realistic secondary spins, we now turn to its gravitational wave signatures. The main question is whether the dynamical differences between regular and chaotic trajectories, as revealed by the Poincar\'e sections, rotation curves, and Lyapunov indicators, leave identifiable imprints on the emitted radiation. Since our goal here is to characterize these waveform features at the phenomenological level rather than to construct fully self-consistent perturbative templates, we adopt the numerical kludge (NK) approach \cite{babak2007a,destounis2021a,cui2025a}.

In the numerical kludge framework, the orbital motion of the spinning secondary is obtained by numerically integrating the MPD equations in Schwarzschild spacetime, while the gravitational radiation is generated from the same relativistic trajectory through an approximate flat-space prescription \cite{babak2007a}. In this way, the strong-field dynamics of the source are retained at the orbital level, whereas the waveform construction remains computationally efficient. Although approximate, this method has been shown to reproduce the main qualitative features of EMRI signals and is therefore well suited for the present exploratory study of chaotic imprints in gravitational waves \cite{babak2007a,destounis2021a}.

In the present analysis, we work within the leading quadrupole approximation of the NK scheme. The mass quadrupole moment of the source is written as
\begin{equation}
I^{ij}(t')=\mu\,Z^i(t')Z^j(t'),
\label{eq:nk_quadrupole}
\end{equation}
where \(Z^i\) denotes the pseudo-flat-space trajectory of the secondary, and \(t'=t-|\mathbf{x}-\mathbf{x}'|\) is the retarded time, with \(\mathbf{x}\) and \(\mathbf{x}'\) denoting the observer and source positions, respectively. The corresponding waveform tensor is then obtained from the second time derivative of the quadrupole moment,
\begin{equation}
h^{ij}(t,\mathbf{x})=\frac{2}{D}\,\ddot{I}^{ij}(t')\big|_{t'=t-|\mathbf{x}-\mathbf{x}'|},
\label{eq:nk_hij}
\end{equation}
where \(D\) is the source-observer distance. The two gravitational wave polarizations \(h_+(t)\) and \(h_\times(t)\) are obtained by projection onto the polarization basis \cite{babak2007a,destounis2021a}. For notational simplicity, we suppress the distinction between \(t\) and \(t'\) below.

Since our aim is to characterize differences between regular and chaotic signals after detector response, we do not display the polarizations \(h_+\) and \(h_\times\) separately in the main results. Instead, we work directly with the detector strain. In the time domain, this is written as
\begin{equation}
h(t)=F_+\,h_+(t)+F_\times\,h_\times(t),
\label{eq:detector_time}
\end{equation}
where \(F_{+,\times}\) are the effective antenna pattern factors for the fixed detector configuration adopted here. In the frequency domain, we use the corresponding approximate detector response in schematic form,
\begin{equation}
\tilde{h}(f)=\mathcal{D}_+(f;\Omega)\,\tilde{h}_+(f)+\mathcal{D}_\times(f;\Omega)\,\tilde{h}_\times(f),
\label{eq:detector_freq}
\end{equation}
where \(\mathcal{D}_{+,\times}(f;\Omega)\) encode the approximate frequency-domain detector response adopted in our implementation. More specifically, Eq.~\eqref{eq:detector_freq} is constructed for a stationary equal-arm detector, with orbital modulation neglected and finite-arm transfer effects absorbed into \(\mathcal{D}_{+,\times}(f;\Omega)\). Thus, Eq.~\eqref{eq:detector_freq} should be understood as an approximate frequency-domain counterpart of Eq.~\eqref{eq:detector_time}, rather than its exact Fourier transform. The explicit expressions adopted in our calculations are given in Appendix~\ref{app:gw}. Throughout this section, the same fixed detector configuration is used, so that any differences in the detector strains arise solely from the underlying orbital dynamics \cite{destounis2021a,cui2025a}.
\begin{figure}[htbp]
    \centering
    \includegraphics[width=\columnwidth]{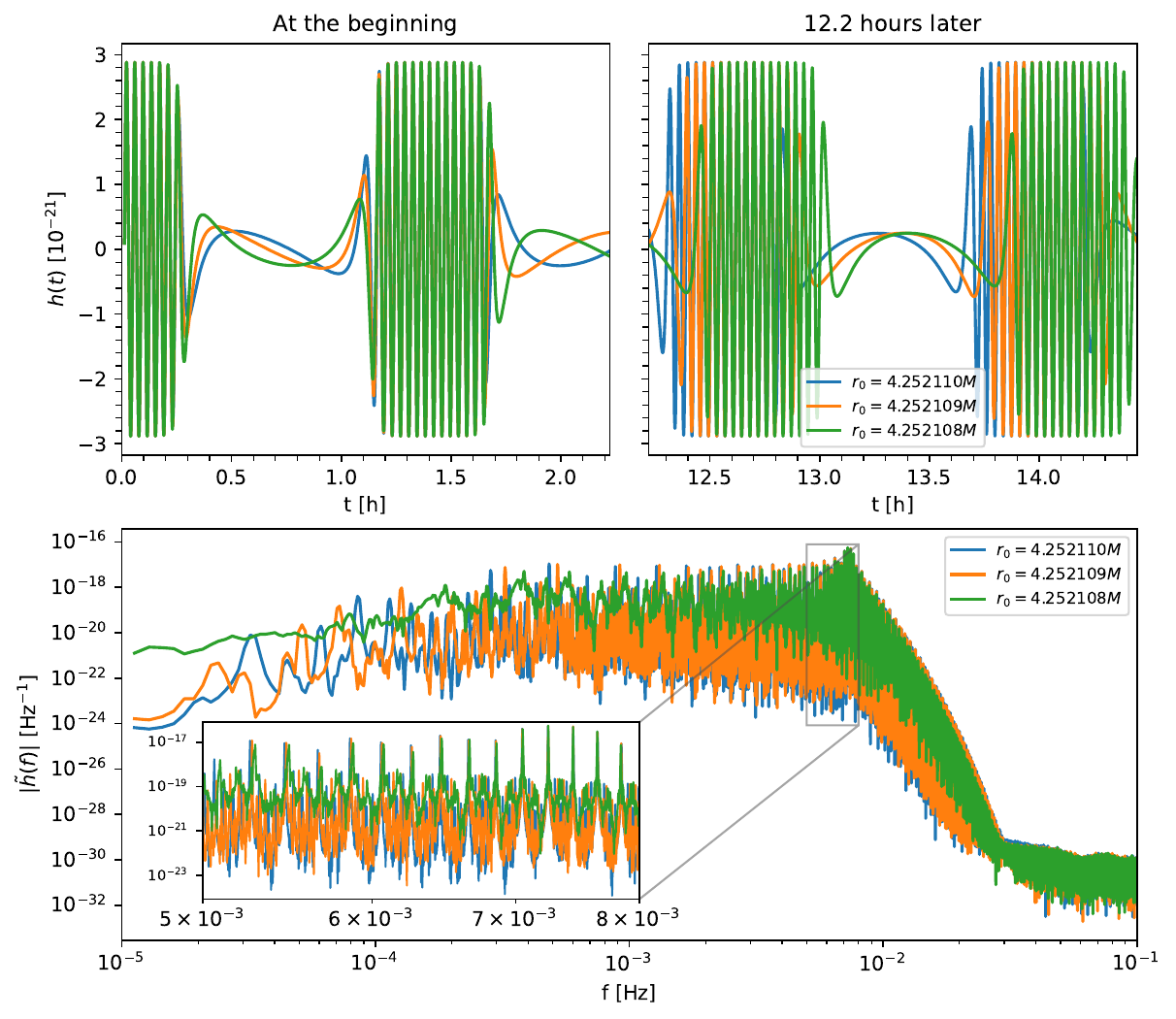}
    \caption{Top panels: Detector waveforms \(h(t)\) for three trajectories from Fig.~\ref{fig:Poincare} with \(S=10^{-5}\mu M\). The green line shows the chaotic orbit with \(r_0=4.252108M\), while the orange and blue lines show the neighboring regular orbits with \(r_0=4.252109M\) and \(r_0=4.252110M\), respectively. The right panel shows the corresponding waveforms \(12.2\) hours later. Bottom panel: Corresponding spectra of the detector strain \(|\tilde{h}(f)|\). The inset shows a zoomed-in view of the frequency interval \(0.005\text{--}0.008~\mathrm{Hz}\).
    }
    \label{fig:h_time_freq}
\end{figure}

The corresponding detector waveforms and their frequency spectra are shown in Fig.~\ref{fig:h_time_freq}. From the top panels, one sees that in the time domain it is difficult to distinguish the chaotic signal from the neighboring regular ones by simple visual inspection alone. At the beginning of the evolution, the three waveforms are very similar, and even after \(12.2\) hours the chaotic waveform only develops a somewhat larger phase offset relative to the regular signals, without exhibiting an immediately obvious qualitative difference. In this sense, the time-domain morphology by itself does not provide a particularly sharp discriminator between chaotic and regular motion for these nearby trajectories.

The distinction becomes much clearer in the frequency domain. As shown in the bottom panel of Fig.~\ref{fig:h_time_freq}, the dominant peaks of the chaotic orbit remain broadly aligned with those of the neighboring regular orbits, indicating that the strongest emission is still concentrated around similar characteristic frequencies. However, compared with the regular cases, the chaotic spectrum is much more populated between the principal peaks, with numerous small spikes filling the inter-peak region and producing a more continuous spectral distribution. By contrast, the regular signals retain a relatively discrete harmonic structure. Such a loss of spectral discreteness is a distinctive signature of the underlying chaotic dynamics and is consistent with previous studies of gravitational radiation from chaotic orbital motion \cite{suzuki1999a}. The nearly flat plateau at high frequencies appears when the spectrum falls to the level of double-precision numerical roundoff error. To make this spectral spreading more explicit, we next examine the gravitational wave energy spectra. For this purpose, we adopt the total energy flux
\begin{equation}
\dot{E}
=
\frac{1}{5}\,\left\langle I^{(3)}_{ij} I^{(3)}_{ij}\right\rangle,
\label{eq:energy_flux_total}
\end{equation}
where \(I^{(3)}_{ij}\equiv d^3 I_{ij}/dt^3\), and \(\langle\cdots\rangle\) denotes averaging over several orbital periods.

\begin{figure}[htbp]
    \centering
    \includegraphics[width=\columnwidth]{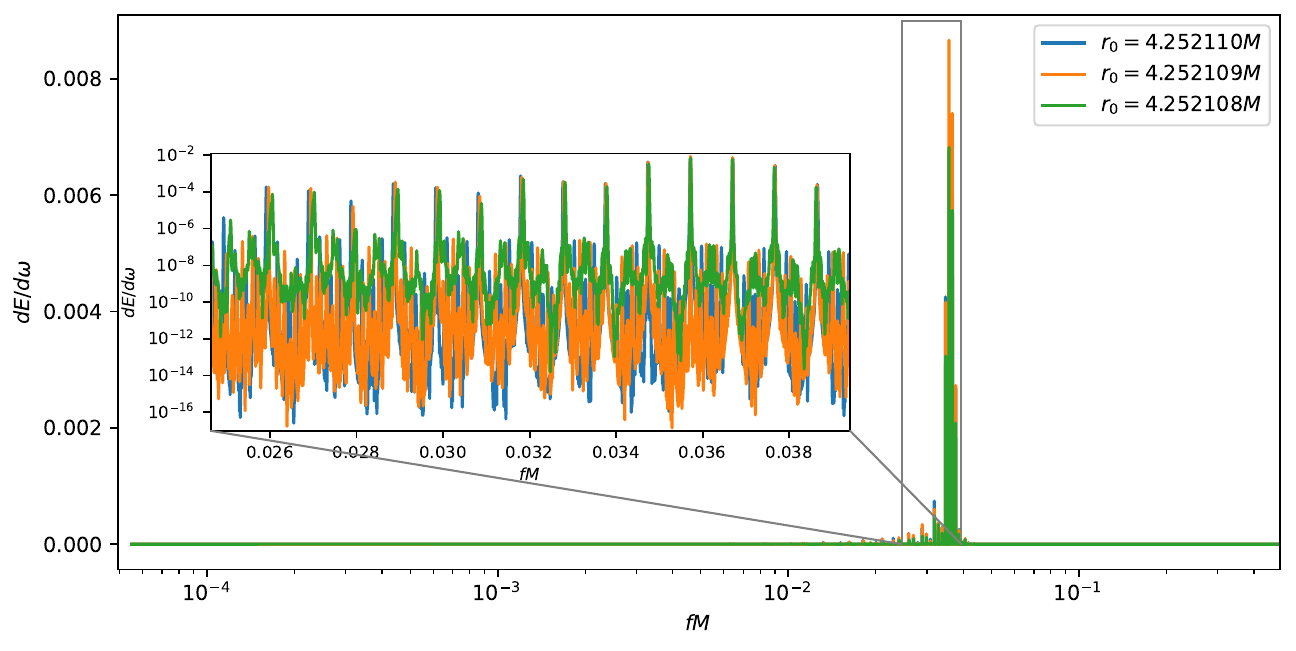}
    \caption{The energy spectra corresponding to the waveforms shown in Fig.~\ref{fig:h_time_freq}. The inset shows a zoomed-in view over the frequency interval \(fM\simeq 0.0246\text{--}0.0394\) \((0.005\text{--}0.008~\mathrm{Hz})\).
    }
    \label{fig:energy_spectrum}
\end{figure}

The corresponding energy spectra are shown in Fig.~\ref{fig:energy_spectrum}. In all three cases, the strongest emission is concentrated in the mHz band, which is also the characteristic sensitivity window of space-based EMRI detectors. The dominant peak of the chaotic orbit still appears in approximately the same frequency range as those of the neighboring regular orbits, indicating that the leading radiative timescales of these nearby trajectories remain similar. However, away from the main peak, the chaotic spectrum exhibits a much richer substructure. 

As shown more clearly in the inset, the regular spectra are still dominated by relatively isolated and sparse lines, whereas the chaotic spectrum contains many more small peaks distributed throughout the interval. As a result, the emitted energy of the chaotic orbit is spread over a substantially broader range of frequencies, giving rise to a more continuous spectral appearance. Such a loss of spectral discreteness is a characteristic signature of chaotic dynamics, reflecting the breakdown of regular periodic or quasiperiodic motion \cite{lichtenberg1992}. This behavior is in qualitative agreement with the results of Suzuki \textit{et al.}, who likewise found that chaotic orbital motion produces significantly more continuous gravitational wave energy spectra than regular motion \cite{suzuki1999a}. More importantly, our results show that this characteristic spectral signature of chaos is retained after projection onto the detector response and is therefore directly reflected in the detector waveforms and spectra studied here.

\subsection{Spectral Flatness as a Quantitative Diagnostic of Chaos}

The analyses in the previous subsection show that, for nearby trajectories with almost identical initial conditions, the detector spectrum of the chaotic orbit is visibly more populated in the inter-peak region and exhibits a more continuous appearance than the spectra of the neighboring regular orbits. To quantify this loss of spectral discreteness, we now introduce spectral flatness as a frequency-domain diagnostic. In gravitational wave data analysis, spectral flatness has previously been used mainly as a measure of spectral whiteness \cite{cuoco2001a}. Here, instead, we use it to characterize the local degree of discreteness or continuity of the gravitational wave power spectrum.

The spectral flatness measure is defined as
\begin{equation}
\xi =
\frac{
\exp\!\left[
\frac{1}{N_s}\int_{-N_s/2}^{N_s/2}\ln P(f)\,df
\right]
}{
\frac{1}{N_s}\int_{-N_s/2}^{N_s/2}P(f)\,df
},
\label{eq:flatness_def}
\end{equation}
where \(P(f)\) is the power spectral density \cite{cuoco2001a}. Since a single flatness measure over the entire frequency band would be dominated by the overall spectral envelope and the strongest peak, we instead adopt a \emph{windowed spectral flatness}. More specifically, for the detector strain \(\tilde{h}(f)\), we define the local power spectrum.
\begin{equation}
P_k \equiv |\tilde{h}(f_k)|^2
\label{eq:flatness_power}
\end{equation}
as the discrete power spectrum within a sliding frequency window \(W(f_c)\) centered at \(f_c\). The corresponding local spectral flatness is then written as
\begin{equation}
\mathcal{F}(f_c)
=
\frac{
\exp\!\left[
\frac{1}{N_w}\sum\limits_{k\in W(f_c)} \ln P_k
\right]
}{
\frac{1}{N_w}\sum\limits_{k\in W(f_c)} P_k
},
\label{eq:local_flatness}
\end{equation}
where \(N_w\) is the number of frequency bins in the window. By construction, \(\mathcal{F}(f_c)\) approaches \(0\) when the spectrum is dominated by sparse, sharp line-like peaks and approaches \(1\) when the spectral power is distributed more uniformly within the same window. Therefore, smaller values of \(\mathcal{F}(f_c)\) correspond to a more discrete and sharply peaked local spectrum, whereas larger values indicate a flatter and more continuous spectral distribution. For convenience, we also define the mean local spectral flatness \(\overline{\mathcal{F}}\) as the average of \(\mathcal{F}(f_c)\) over the selected frequency interval.

\begin{figure}[htbp]
    \centering
    \includegraphics[width=\columnwidth]{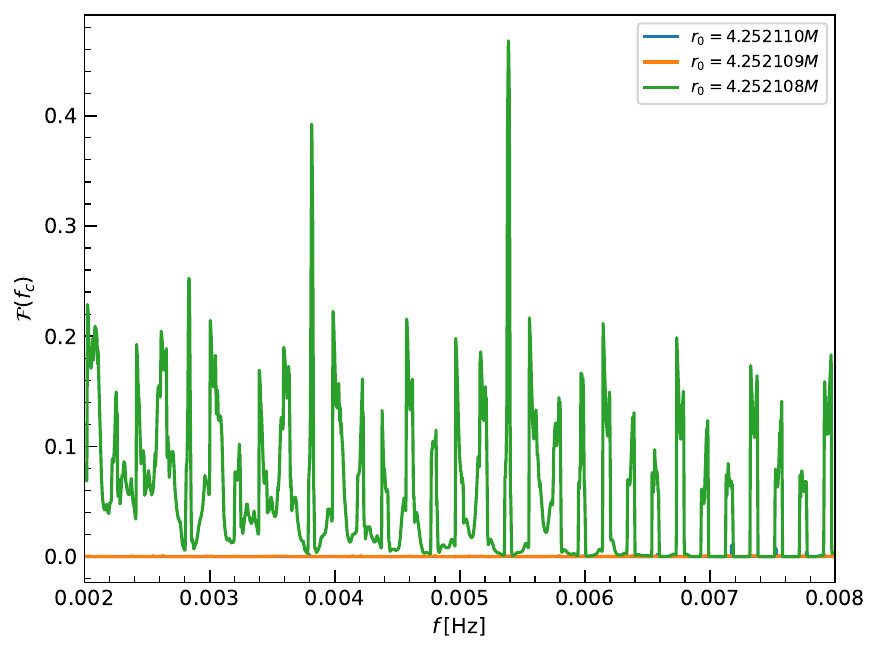}
    \caption{Local spectral flatness \(\mathcal{F}(f_c)\) for the gravitational wave power spectrum of the same trajectories shown in Fig.~\ref{fig:h_time_freq}. The green line denotes the chaotic orbit with \(r_0=4.252108M\), while the orange and blue lines denote the neighboring regular orbits with \(r_0=4.252109M\) and \(r_0=4.252110M\), respectively. The quantity is computed for the LISA detector from the local power spectrum \(P(f)\) over \(0.002\text{--}0.008~\mathrm{Hz}\), using a sliding window of \(81\) frequency bins.
    }
    \label{fig:flatness}
\end{figure}

Figure~\ref{fig:flatness} shows the resulting local spectral flatness for the same three nearby trajectories discussed in the previous subsection. A clear separation between the chaotic and regular cases is immediately visible. Throughout the frequency interval shown, the two regular orbits remain very close to \(\mathcal{F}(f_c)\simeq 0\), indicating that their detector strain spectra are still dominated by sparse and sharply peaked line-like structures within each local window. By contrast, the chaotic orbit exhibits systematically larger values of \(\mathcal{F}(f_c)\), together with pronounced fluctuations across the whole band. This demonstrates that, at the local level, the chaotic detector spectrum is much less discrete and substantially more filled in than the neighboring regular spectra.

\begin{table}[htbp]
\centering
\caption{Mean local spectral flatness \(\overline{\mathcal{F}}\) for the three trajectories shown in Fig.~\ref{fig:flatness}.}
\label{tab:flatness_summary}
\renewcommand{\arraystretch}{1.15}
\begin{tabular*}{0.8\columnwidth}{@{\extracolsep{\fill}}cc@{}}
\hline\hline
Trajectory & $\overline{\mathcal{F}}$ \\
\hline
$r_0=4.252110M$ & $7.42\times10^{-5}$ \\
$r_0=4.252109M$ & $1.25\times10^{-4}$ \\
$r_0=4.252108M$ & $5.43\times10^{-2}$ \\
\hline\hline
\end{tabular*}
\end{table}

This trend is also reflected quantitatively in Table~\ref{tab:flatness_summary}. The chaotic orbit yields a mean local spectral flatness that is several hundred times larger than that of the two neighboring regular orbits. At the same time, its value remains far below the maximally flat limit \(\overline{\mathcal F}=1\). Therefore, although the chaotic spectrum is much more continuous than the regular line-like spectra, it remains far from featureless white noise and still preserves a well-defined spectral structure; in particular, an approximately equally spaced modulation pattern is also visible in the local spectral flatness.

This provides a quantitative confirmation of the qualitative picture already seen in the spectra of the detector strain and energy spectra. In the regular cases, most of the spectral power is concentrated in a small number of isolated frequency bins, so the geometric mean within each window is strongly suppressed and the corresponding flatness remains close to zero. For the chaotic orbit, however, the inter-peak region is populated by many additional small spectral components, so that the power within a given window is distributed more broadly, and the local flatness increases accordingly. In this sense, the loss of spectral discreteness discussed in the previous subsection is not merely a visual impression but can be captured by a simple and consistent frequency-domain statistic. Since \(\mathcal{F}(f_c)\) is computed here from the detector strain rather than from the source-frame polarizations, Fig.~\ref{fig:flatness} and Table~\ref{tab:flatness_summary} further show that this characteristic spectral signature of chaos survives the detector response.

Taken together, Fig.~\ref{fig:flatness} and Table~\ref{tab:flatness_summary} show that the spectral distinction between chaotic and regular trajectories can be quantified in a simple way. The chaotic signal exhibits a systematically larger local spectral flatness than the neighboring regular signals while still remaining far below the white-noise limit. This confirms that the loss of spectral discreteness is a consistent frequency-domain signature of chaos that survives the detector response. We next examine how this difference is reflected in detector-relevant quantities such as the characteristic strain, the signal-to-noise ratio, and the overlap.

\subsection{Characteristic strains, signal-to-noise ratios, and overlap under spin variations}

We now turn to the impact of varying the secondary spin on the gravitational wave signal. As already shown by the phase-space analysis in Fig.~\ref{fig:Poincare}, even relatively small changes in the spin can correspond to qualitatively different orbital structures. In particular, the three cases considered here do not merely represent nearby perturbations of the same phase-space trajectory family; rather, their underlying phase-space portraits are already markedly different, with the \(S=1.8\times10^{-5}\mu M\) case lying inside the chaotic region, while the neighboring smaller-spin cases remain regular. It is therefore natural to ask how such dynamical differences are reflected in the observable waveform morphology.

To address this question, we first introduce the characteristic strain of the detector-contracted signal,
\begin{equation}
h_c(f)=2f\,|\tilde{h}(f)|,
\label{eq:hc_def}
\end{equation}
and the corresponding noise amplitude
\begin{equation}
h_n(f)=\sqrt{f\,P_n(f)},
\label{eq:hn_def}
\end{equation}
where \(P_n(f)\) is the one-sided detector noise spectral density, including the Galactic confusion noise. The associated signal-to-noise ratio (SNR) \(\rho\) is defined as
\begin{equation}
\rho^2=4\int_0^\infty \frac{|\tilde{h}(f)|^2}{P_n(f)}\,df.
\label{eq:snr_def}
\end{equation}
In the numerical implementation, we further employ a long-wavelength approximation to estimate the corresponding sky-averaged SNR for the detector configuration adopted here.

\begin{figure}[htbp]
    \centering
    \includegraphics[width=\columnwidth]{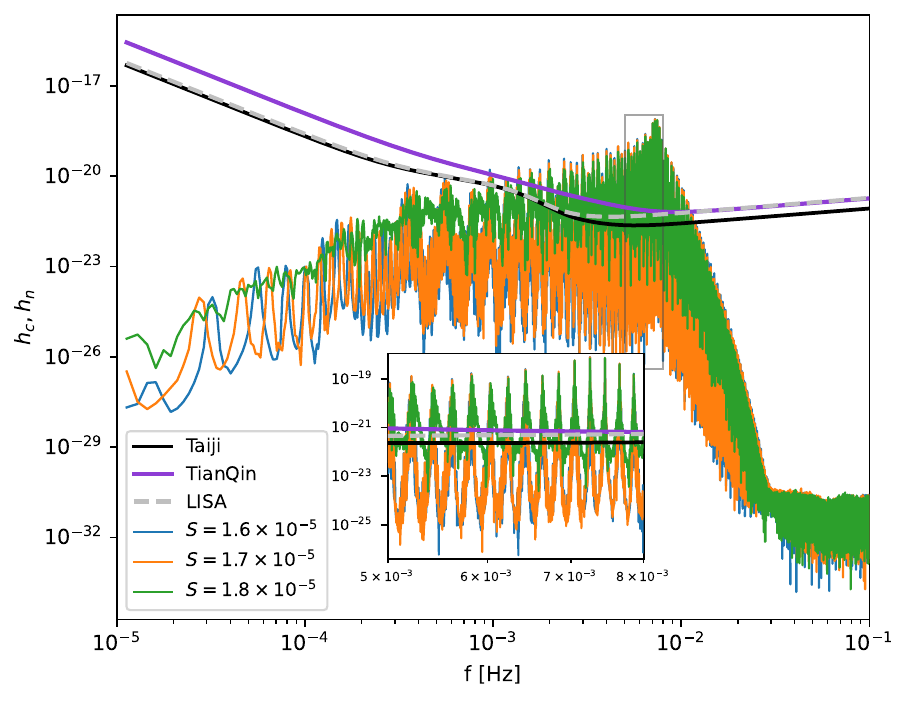}
    \caption{The characteristic strains \(h_c(f)\) for three trajectories with the same initial radius \(r_0=4.252117M\) but different secondary spins: \(S=1.6\times10^{-5}\mu M\) (blue), \(S=1.7\times10^{-5}\mu M\) (orange), and \(S=1.8\times10^{-5}\mu M\) (green), where the last case corresponds to a chaotic orbit. Here \(h_c(f)\) is computed from the LISA-contracted signal. The black, purple, and light-gray dashed curves show the noise amplitudes \(h_n(f)\) of Taiji, TianQin, and LISA, respectively. The inset shows a zoomed-in view of the frequency interval \(0.005\text{--}0.008~\mathrm{Hz}\).}
    \label{fig:hc_spin}
\end{figure}

The resulting characteristic strains are shown in Fig.~\ref{fig:hc_spin}. In all three cases, the signal power is concentrated in the mHz band, which overlaps well with the most sensitive frequency window of the planned space-based detectors. Frequencies at which the characteristic strain lies above the corresponding noise amplitude mark the bands where the signals are expected to be observable. At the same time, the three spectra exhibit substantial differences in their detailed fine structure. In particular, the chaotic case with \(S=1.8\times10^{-5}\mu M\) displays a much denser and more continuous distribution of spectral peaks than the two regular cases, in agreement with the frequency-domain and spectral-flatness analyses presented above.

The corresponding SNRs are summarized in Table~\ref{tab:snr_spin}. One can see that all three signals yield comparable SNRs, with \(\rho_{\mathrm{LISA}}\simeq 43\), \(\rho_{\mathrm{Taiji}}\simeq 90\), and \(\rho_{\mathrm{TianQin}}\simeq 35\). Thus, the large waveform differences discussed below are not due to any dramatic suppression or enhancement of the overall signal strength. Rather, they originate from the qualitative change in the underlying orbital dynamics induced by the spin variation.

\begin{table}[htbp]
\centering
\caption{SNR for the three trajectories shown in Fig.~\ref{fig:hc_spin}.}
\label{tab:snr_spin}
\renewcommand{\arraystretch}{1.15}
\begin{tabular*}{\columnwidth}{@{\extracolsep{\fill}}cccc@{}}
\hline\hline
Spin & LISA & Taiji & TianQin \\
\hline
$S=1.6\times10^{-5}\mu M$ & 43.475 & 89.534 & 34.640 \\
$S=1.7\times10^{-5}\mu M$ & 43.525 & 89.642 & 34.687 \\
$S=1.8\times10^{-5}\mu M$ & 43.708 & 90.034 & 34.856 \\
\hline\hline
\end{tabular*}
\end{table}

To quantify the waveform difference more directly, we further consider the normalized frequency-domain overlap between two detector strains,
\begin{equation}
\mathcal{O}(h_1,h_2)
=
\frac{\langle h_1|h_2\rangle}
{\sqrt{\langle h_1|h_1\rangle\langle h_2|h_2\rangle}},
\label{eq:overlap_def}
\end{equation}
where the noise-weighted inner product is defined by
\begin{equation}
\langle h_1|h_2\rangle
=
4\,\mathrm{Re}\int_0^\infty
\frac{\tilde{h}_1(f)\tilde{h}_2^*(f)}{P_n(f)}\,df.
\label{eq:inner_product_def}
\end{equation}
A smaller overlap indicates a stronger difference between the two signals in the detector band.

For the two neighboring regular trajectories, \(S=1.6\times10^{-5}\mu M\) and \(S=1.7\times10^{-5}\mu M\), the overlap is still relatively large, \(\mathcal{O}=0.678\). By contrast, when the spin is increased only slightly further from \(S=1.7\times10^{-5}\mu M\) to \(S=1.8\times10^{-5}\mu M\), corresponding to a change of merely \(1\%\) of the astrophysical spin limit considered here, the orbit enters the chaotic region and the overlap drops sharply to \(\mathcal{O}=0.159\). Thus, even such a small spin variation can produce a dramatic waveform difference once it drives the system across the boundary from regular to chaotic motion, despite the fact that the corresponding signals have very similar overall SNRs and lie in the same detector sensitivity band.

Taken together, the analyses presented in Sec.~IV show that chaotic orbits possess clear and distinctive gravitational-wave signatures. Their detector strains retain the same dominant peaks as those of neighboring regular orbits, indicating that the strongest emission is still concentrated around similar characteristic frequencies, while the spectral region between these peaks becomes much more densely populated and hence more continuous. This difference is further confirmed by the larger local spectral flatness of the chaotic signal, which nevertheless remains far below the white-noise limit.

At the same time, the characteristic strains lie in the mHz band relevant for space-based detectors and yield comparable overall signal-to-noise ratios. More importantly, even a very small change in the secondary spin can produce a dramatic drop in the overlap once the system crosses from regular to chaotic motion. In this sense, chaotic dynamics leave a characteristic spectral imprint that survives the detector response and may provide a useful handle for identifying such behavior in future gravitational-wave data analysis.

\section{Discussion and Conclusions}

In this work, we have investigated the chaotic dynamics of a spinning secondary in a Schwarzschild EMRI system within the pole-dipole approximation, with particular emphasis on two questions: whether chaos can arise for astrophysically realistic secondary spins, and whether such chaos leaves identifiable imprints on the corresponding gravitational-wave signal. Using the geodesic homoclinic orbit as a reference configuration, we analyzed the nearby phase-space structure through Poincar\'e sections, rotation curves, and Lyapunov indicators, and then computed the associated gravitational wave signals and detector response within a numerical kludge framework.

We find that, for the EMRI mass ratio \(q\equiv \mu/M=10^{-4}\) considered in this work, chaotic motion persists from \(0.1\) times the astrophysical spin limit, \(S=10^{-5}\mu M\), up to the full limit, \(S=10^{-4}\mu M\). We show that chaos is already present at \(S=10^{-5}\mu M\) and becomes increasingly prominent as the spin approaches the astrophysical upper bound. This provides strong numerical evidence that spin-induced chaos in Schwarzschild EMRIs is not confined to unrealistically large spins but already exists for astrophysically realistic secondary spins. More importantly, our numerical results suggest that second-order spin effects may inevitably lead to chaos even for astrophysically realizable secondary spins, a possibility that deserves to be established on firmer theoretical grounds in future work. We also find a sharp jump in the rotation curve near \(\nu_\theta \simeq 3/4\), which our analysis suggests is associated with a resonant separatrix structure induced by the secondary spin near the \(3/4\) resonance. This indicates that spin can trigger abrupt changes in the rotation number, or equivalently in the underlying frequency ratio, with possible implications for the associated gravitational wave signal.

These dynamical differences are reflected directly in the gravitational-wave signal. For nearby trajectories with almost identical initial conditions, the detector waveforms remain rather similar in the time domain and are not easily distinguished by visual inspection alone. In the frequency domain, however, the distinction becomes much clearer: compared with neighboring regular orbits, the chaotic orbit exhibits a much denser population of small spectral components between the principal peaks, giving rise to a broader and less discrete spectral appearance. The same trend is visible in the energy spectra. The chaos-induced spectral flattening is further quantified by the local spectral flatness, whose mean value for the chaotic orbit is several hundred times larger than that for the neighboring regular orbits, while still remaining far below the white-noise limit. Moreover, the corresponding characteristic strains lie in the mHz band relevant for space-based detectors and yield comparable overall signal-to-noise ratios, whereas even a very small change in the secondary spin can cause a dramatic drop in the overlap once the system crosses from regular to chaotic motion. Taken together, these results show that the chaos-induced loss of spectral discreteness remains visible after detector response and may provide a distinctive detector-level signature in future space-based observations.

Several directions merit further investigation. A natural next step is to extend the present analysis to Kerr spacetime. Although its phase space is higher-dimensional and therefore harder to visualize directly, the framework developed here for locating the onset of chaos should remain applicable, raising the prospect that chaotic orbits associated with astrophysically realizable secondary spins may also be identified in Kerr EMRIs and leave detectable gravitational wave signatures. More broadly, chaos in EMRIs may arise not only from the secondary spin considered here, but also from broken background symmetries or environmental perturbations \cite{fang2019,hua2026,destounis2021a}, so distinguishing these different origins of nonintegrability at the level of orbital dynamics and gravitational wave observables will be an important task, especially once the relevant parameter space becomes too high-dimensional for direct exploration. Data-driven methods may then offer an efficient way to map or classify chaotic regions, extending recent machine-learning applications to EMRI detection and gravitational-wave signal reconstruction \cite{zhao2024,lai2025}. Finally, the waveform analysis presented here is based on a numerical kludge description and neglects radiation reaction. It will therefore be important to combine more accurate waveform calculations with dissipative inspiral effects, in order to determine whether the chaotic spectral features found here can persist long enough to leave observable imprints in realistic inspirals crossing the chaotic region.

\section*{Acknowledgements}

The authors would like to warmly thank Ond\v{r}ej Zelenka, Wen-Biao Han, and Hongbao Zhang for valuable discussions and advice. We also thank Zhen-Tao He, Zhutong Hua, and Jia Du for helpful exchanges and discussions. This research is supported in part by the National Key R\&D Program of China, grant number 2020YFC2201300, and the National Natural Science Foundation of China, grant numbers 12035016, 12375058 and 12361141825.

\appendix

\section{Numerical integration of the MPD equations}
\label{app:mpd}

We evolve the full MPD system in the original variables \((x^\mu,P^\mu,S^{\mu\nu})\). Once the TD supplementary condition is imposed, the four velocity \(v^\mu\) is reconstructed from \(P^\mu\), \(S^{\mu\nu}\), and the background curvature, so that the equations of motion form a closed 14-dimensional first order system.

For each orbit, we specify the constants \((E,J_z,S,\mu)\) together with an initial pair \((r_0,P_{r,0})\), and in all calculations reported here we set \(P_{r,0}=0\). In all numerical examples considered in this work, the initial point is chosen on the equatorial plane
\begin{equation}
\theta_0=\frac{\pi}{2}, \qquad \phi_0=0,
\label{eq:app_ic_plane}
\end{equation}
and the spatial axes are aligned with the total angular momentum such that
\begin{equation}
J_x=J_y=0, \qquad J_z=J .
\label{eq:app_ic_J}
\end{equation}
Under these conditions, the independent components of the spin tensor are
\begin{equation}
\begin{gathered}
S^{\theta\phi}=0,\qquad
S^{r\theta}=-\frac{P_\theta}{r},\qquad
S^{\phi r}=\frac{P_\phi-J_z}{r},\\[3pt]
S^{tr}=-\frac{P_\theta^2+P_\phi(P_\phi-J_z)}{rP_t},\qquad
S^{t\theta}=\frac{P_rP_\theta}{rP_t},\\[3pt]
S^{t\phi}=\frac{P_r(P_\phi-J_z)}{rP_t}.
\end{gathered}
\label{eq:app_spin_components}
\end{equation}
Therefore, once \((E,J_z,S,\mu,r_0,P_{r,0})\) are fixed, the remaining unknown momentum components
\((P_t,P_\theta,P_\phi)\) are determined from the three constraint equations
\begin{subequations}
\label{eq:app_ic_constraints}
\begin{align}
0 &= E + P_t - \frac{M}{r^2}S^{tr}, 
\label{eq:app_ic_constraints_a}\\
0 &= \mu^2 - \frac{P_t^2}{f(r)} + f(r)P_r^2
      + \frac{P_\theta^2}{r^2} + \frac{P_\phi^2}{r^2},
\label{eq:app_ic_constraints_b}\\
0 &= S^2 - \frac{1}{2}S_{\mu\nu}S^{\mu\nu},
\label{eq:app_ic_constraints_c}
\end{align}
\end{subequations}

with the expressions in Eq.~\eqref{eq:app_spin_components} substituted. These nonlinear equations are solved using the Newton-Raphson method. In practice, the initial condition solve is performed in 128-bit \texttt{BigFloat} arithmetic to suppress the loss of significance associated with small differences such as \(E+P_t\) and \(J_z-P_\phi\). The converged initial data are then converted to standard double precision for the subsequent orbital evolution.

The motivation for this choice is mainly practical. With \(P_{r,0}=0\), varying \(r_0\) amounts to scanning the initial conditions along a simple line in the \((r,P_r)\) plane of the Poincar\'e section. In regular regions, this yields a particularly clean and systematic sampling of neighboring invariant curves and makes the resulting phase-space portrait easier to organize and compare. It is also convenient for resolving the sharp structures near the tip of the main island and near the associated separatrix. By contrast, if the initial data are parametrized by a spin-tilt angle, as in the construction used by Suzuki and Maeda \cite{suzuki1997a}, \(P_r\) is not fixed a priori, and the corresponding scan on the Poincar\'e section is less direct.

For fixed \((E,J_z,S,\mu)\), the choice \(P_{r,0}=0\) does not leave \(r_0\) arbitrary. Real initial data exist only for those radii at which the constraint equations admit real solutions for \((P_t,P_\theta,P_\phi)\). On the equatorial plane, this is equivalent to requiring that the prescribed energy intersects the effective-potential band
\begin{equation}
E = V_{\mathrm{eff}}^{(\pm)}\!\left(r,\frac{\pi}{2};J,S\right),
\label{eq:app_r0_band}
\end{equation}
so that the admissible interval of initial radii is bounded by the corresponding intersection points. These are the blue-violet solid lines indicated in the inset of Fig.~\ref{fig:Veff}.

For the time evolution, we use the fixed-step \texttt{IRKGL16} integrator, i.e., a 16th-order Gauss-Legendre collocation method, implemented through the Julia package \texttt{IRKGaussLegendre.jl} \cite{IRKGaussLegendre_jl}. This class of symmetric structure-preserving integrators is particularly well suited to long-term integrations of relativistic orbital dynamics, as it provides substantially better conservation of the first integrals than standard explicit schemes over the integration times relevant here \cite{seyrich2012b,seyrich2013}. Throughout this work, we adopt a step size \(\Delta \tau = 1\,M \).

During the evolution, we monitor the conserved quantities \(E\), \(J_z\), \(\mu^2\), and \(S^2\) as diagnostics of numerical accuracy. For the ordinary orbital integrations reported in the main text, we use standard double-precision floating-point arithmetic, for which the relative errors of the conserved quantities remain below \(10^{-14}\) up to \(\tau=10^6 M\). For the FLI calculations, where the long-time tracking of nearby trajectories requires higher numerical accuracy, we instead use quadruple-precision arithmetic; in this case, the relative errors of the conserved quantities remain below \(10^{-30}\) up to \(\tau=10^6 M\).

\section{Gravitational wave generation and detector response}
\label{app:gw}

\begin{figure*}[t]
    \centering
    \includegraphics[width=\textwidth]{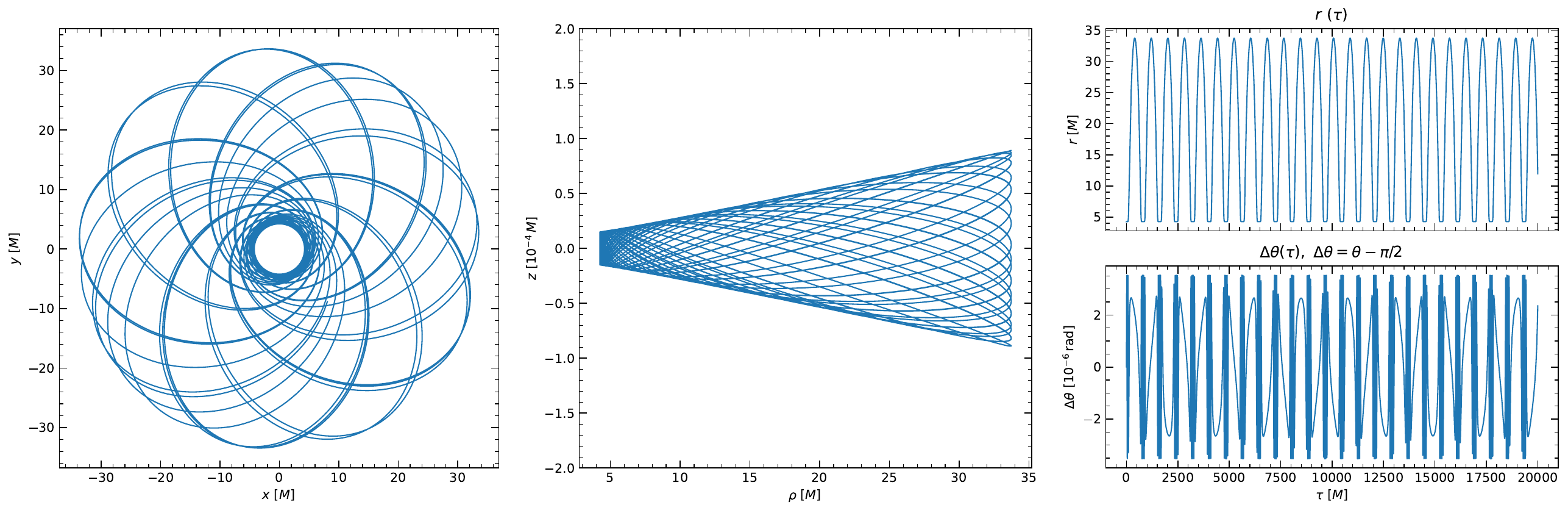}
    \caption{Representative bound orbit with $S=10^{-5}\mu M$ and $r_0=4.252109M$, corresponding to the regular trajectory shown by the orange curve in Fig.~\ref{fig:h_time_freq}. The left panel shows the orbital trajectory projected onto the $(x,y)$ plane, while the middle panel displays the meridional projection in the $(\rho,z)$ plane, with $\rho=\sqrt{x^2+y^2}$. The right panels show the corresponding evolution of the radial coordinate $r(\tau)$ and the polar deviation $\Delta\theta(\tau)\equiv\theta(\tau)-\pi/2$ as functions of proper time. The motion is characterized by a large radial oscillation together with a much weaker out-of-equatorial modulation.
    }
    \label{fig:app_orbit_structure}
\end{figure*}

The gravitational-wave signals used in this work are constructed within the numerical kludge (NK) framework. Following the standard NK prescription, the bound trajectory obtained from the MPD evolution is re-interpreted as an effective source trajectory in flat-space spherical coordinates identified with the Schwarzschild coordinates \((r,\theta,\phi)\). The corresponding Cartesian coordinates are
\begin{equation}
x=r\sin\theta\cos\phi,\qquad
y=r\sin\theta\sin\phi,\qquad
z=r\cos\theta .
\label{eq:app_xyz}
\end{equation}
The source quadrupole tensor is then taken as
\begin{equation}
I_{ij}=\mu x_i x_j ,
\label{eq:app_quadrupole}
\end{equation}
with \(\mu\) the mass of the secondary. The waveform polarizations are computed from the second time derivative of the quadrupole moment with respect to the coordinate time \(t\), while the radiated power is obtained from the third time derivative of its symmetric trace-free part.

To define the two source polarizations, we introduce an orthonormal polarization triad \((\hat{\mathbf p},\hat{\mathbf q},\hat{\mathbf n})\), where \(\hat{\mathbf n}\) points along the line of sight and \(\hat{\mathbf p}\), \(\hat{\mathbf q}\) span the plane transverse to the propagation direction. The corresponding polarization tensors are
\begin{equation}
e^{+}_{ij}=p_i p_j-q_i q_j,\qquad
e^{\times}_{ij}=p_i q_j+q_i p_j .
\label{eq:app_pol_tensors}
\end{equation}
The transverse-traceless metric perturbation generated by the quadrupole moment is written as
\begin{equation}
h^{\mathrm{TT}}_{ij}(t,\mathbf{x})
=
\frac{2}{D}\,
\Lambda_{ij,kl}(\hat{\mathbf n})\,
\ddot I_{kl}(t')
\Big|_{\,t'=\,t-|\mathbf{x}-\mathbf{x}'|},
\label{eq:app_hTT_quad}
\end{equation}
where \(D\) is the source-observer distance, \(t'\) is the retarded time, \(\mathbf{x}\) and \(\mathbf{x}'\) denote the observer and source positions, respectively. The transverse-traceless projector is
\begin{equation}
\Lambda_{ij,kl}(\hat{\mathbf n})
=
P_{ik}P_{jl}
-\frac{1}{2}P_{ij}P_{kl},
\qquad
P_{ij}=\delta_{ij}-n_i n_j ,
\label{eq:app_TT_projector}
\end{equation}
with \(\hat{\mathbf n}\) the line-of-sight unit vector from source to observer. The two source polarizations are then obtained by projecting \(h^{\mathrm{TT}}_{ij}\) onto the polarization basis,
\begin{equation}
h_+(t)=\frac{1}{2}e_+^{ij}h^{\mathrm{TT}}_{ij}(t,\mathbf{x}),
\qquad
h_\times(t)=\frac{1}{2}e_\times^{ij}h^{\mathrm{TT}}_{ij}(t,\mathbf{x}).
\label{eq:app_hpm_def}
\end{equation}
In the numerical implementation adopted here, the transverse-traceless projection is evaluated in a polarization basis adapted to the line of sight. For notational simplicity, the distinction between \(t\) and \(t'\) is suppressed below whenever no confusion can arise.

For the detector-level analysis, the source polarizations are passed through the response of a space-based Michelson-type interferometer. In the frequency domain, the detector response can be written as
\begin{equation}
\tilde h(f)=\mathcal{D}^{ij}(f,\hat{\mathbf n})
\left(
e^{+}_{ij}\,\tilde h_+(f)+e^{\times}_{ij}\,\tilde h_\times(f)
\right),
\label{eq:app_detector_freq}
\end{equation}
where the detector tensor is
\begin{equation}
\mathcal{D}^{ij}(f,\hat{\mathbf n})
=
\frac12
\left[
u^i u^j\,\mathcal{T}_u(f,\hat{\mathbf n})
-
v^i v^j\,\mathcal{T}_v(f,\hat{\mathbf n})
\right].
\label{eq:app_detector_tensor}
\end{equation}
Here \(\mathbf{u}\) and \(\mathbf{v}\) denote the two arm directions, and \(\mathcal{T}_u\), \(\mathcal{T}_v\) are the corresponding arm transfer functions.

In the long-wavelength limit, the frequency-domain response of Eq.~\eqref{eq:app_detector_freq} simplifies and its time-domain counterpart becomes
\begin{equation}
h_{\rm det}(t)=F_+ h_+(t)+F_\times h_\times(t),
\label{eq:app_detector_strain}
\end{equation}
with
\begin{equation}
F_+ = D^{ij}e^{+}_{ij},\qquad
F_\times = D^{ij}e^{\times}_{ij},
\label{eq:app_pattern_def}
\end{equation}
where \(D^{ij}=\frac12(u^i u^j-v^i v^j)\). For the arm-angle convention adopted in our implementation the antenna pattern functions become
\begin{equation}
F_+ =
\frac{\sqrt{3}}{2}
\left[
\frac{1+\cos^2\theta}{2}\sin 2\phi \cos 2\psi
+\cos\theta \cos 2\phi \sin 2\psi
\right],
\label{eq:app_Fplus}
\end{equation}
\begin{equation}
F_\times =
\frac{\sqrt{3}}{2}
\left[
\frac{1+\cos^2\theta}{2}\sin 2\phi \sin 2\psi
-\cos\theta \cos 2\phi \cos 2\psi
\right].
\label{eq:app_Fcross}
\end{equation}
These expressions are equivalent to the standard long-wavelength response of a \(60^\circ\) Michelson interferometer, up to a trivial choice of azimuthal origin.

For the waveform calculations presented in this work, we adopt a central black-hole mass \(M=10^6\,M_\odot\), a secondary mass \(\mu=100\,M_\odot\), and a source distance \(D=1\,\mathrm{Gpc}\). For the detector response, we choose the extrinsic angles \(\theta=\pi/4\), \(\phi=\pi/3\), and \(\psi=0\).

To make the origin of the spectral structure discussed in the main text more transparent, we show in Fig.~\ref{fig:app_orbit_structure} the source trajectory underlying the regular orbit with $S=10^{-5}\mu M$ and $r_0=4.252109M$, corresponding to the orange curve in Fig.~\ref{fig:h_time_freq}. The orbit displays a clear zoom-whirl pattern: long zoom segments connect repeated near-pericenter whirls, producing the characteristic rosette-like projection in the orbital plane. At the same time, because the spin is small, the motion remains confined to a thin wedge around the equatorial plane, as discussed in Sec.~II.C. This is also reflected in the right panels of Fig.~\ref{fig:app_orbit_structure}: the radial coordinate \(r(\tau)\) exhibits a strong and nearly periodic oscillation, whereas the polar deviation \(\Delta\theta(\tau)\) stays much smaller and combines rapid oscillations with a slower long-term modulation.

\begin{figure}[htbp]
    \centering
    \includegraphics[width=\columnwidth]{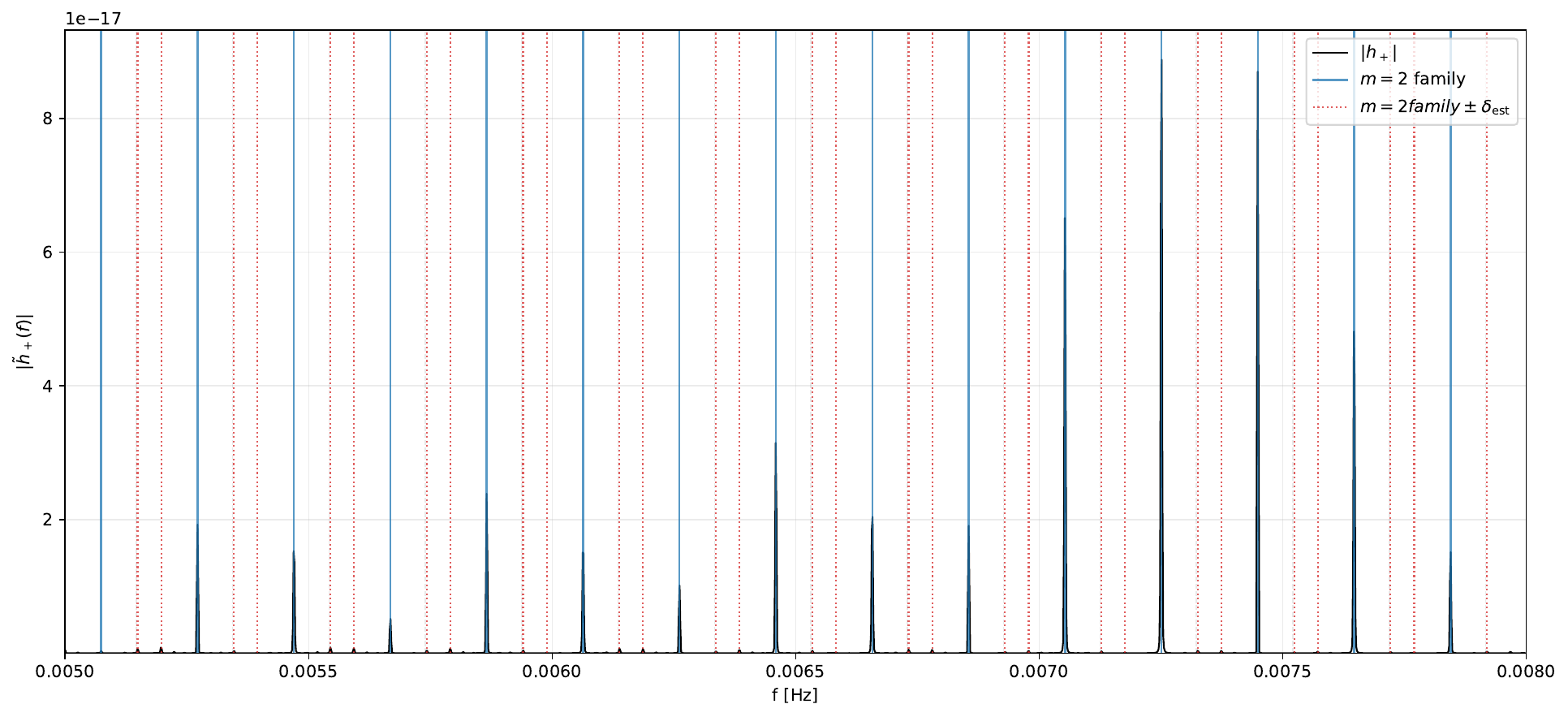}
    \caption{Frequency-domain source spectrum $|\tilde h_+(f)|$ for the orbit with $S=10^{-5}\mu M$ and $r_0=4.252109M$, computed in the numerical kludge model before applying the detector response. The blue full lines mark the dominant $m=2$ family, $f=2f_{\phi,\mathrm{wind}}+n f_r$, which accurately traces the main spectral peaks. The red dotted lines indicate the shifted sideband pattern $f=2f_{\phi,\mathrm{wind}}+n f_r\pm\delta_{\mathrm{est}}$. The figure illustrates a two-layer spectral structure: a primary comb associated with the dominant azimuthal-radial family, together with weaker sidebands around it.
    }
    \label{fig:app_spectrum_structure}
\end{figure}

This orbital structure is directly imprinted on the waveform spectrum. For the present trajectory, the dominant frequencies extracted from the coordinate-time motion are \(f_r\approx1.98\times10^{-4}\,\mathrm{Hz}\) and \(f_{\phi,\mathrm{wind}}\approx1.55\times10^{-3}\,\mathrm{Hz}\). The main peaks in the source spectrum are well described by the \(m=2\) family \(f=2f_{\phi,\mathrm{wind}}+n f_r\), shown by the blue solid lines in Fig.~\ref{fig:app_spectrum_structure}. Physically, this primary comb is set by the radial repetition of the zoom-whirl motion together with the dominant quadrupolar azimuthal harmonic. In addition, because \(2f_{\phi,\mathrm{wind}}/f_r\) is close to, but not exactly equal to, the integer \(16\), successive whirl episodes do not repeat with exactly the same azimuthal phase relative to the radial cycle. This produces a slow beat modulation with frequency \(\delta\approx |2f_{\phi,\mathrm{wind}}-16f_r|\), which in turn generates the weaker sidebands around the main \(m=2\) family, indicated by the red dotted lines \(f=2f_{\phi,\mathrm{wind}}+n f_r\pm\delta_{\mathrm{est}}\). Therefore, Fig.~\ref{fig:app_spectrum_structure} provides a source-level explanation for the frequency-domain pattern seen in Fig.~\ref{fig:h_time_freq}: the detector spectrum shown in the main text is inherited from the same underlying azimuthal-radial frequency structure, with the detector response mainly reweighting the amplitudes rather than changing the basic peak locations.

\clearpage
\bibliographystyle{apsrev4-2}
%

\end{document}